\documentclass[preprint]{JHEP} 

\usepackage{epsfig}

\relax

\def\l{\lambda}

\def\be{\begin{equation}}
\def\ee{\end{equation}}
\def\bs{\begin{subequations}}
\def\es{\end{subequations}}

\def\l{\lambda}

\def\d{\partial}

\def\sp{\;\;\;,\;\;\;}
\def\spa{\;\;\;}
\def\r{\rho}

\def\e{\epsilon}
\def\m{\mu}
\def\n{\nu}
\def\r{\rho}
\def\s{\sigma}
\def\a{\alpha}
\def\b{\beta}
\def\f{{\cal F}}
\def\Z{\mbox{\sf Z\hspace{-3.2mm} Z}}

\newcommand\fverb{\setbox\pippobox=\hbox\bgroup\verb}
\newcommand\fverbdo{\egroup\medskip\noindent%
                        \fbox{\unhbox\pippobox}\ }
\newcommand\fverbit{\egroup\item[\fbox{\unhbox\pippobox}]}
\newbox\pippobox
\title{Supersymmetry and Duality in Field Theory and String Theory}

\author{ Elias Kiritsis\\
Physics Department, University of Crete,
71003 Heraklion, GREECE\\
E-mail: \email{kiritsis@physics.uch.gr}}

\preprint{\hepth{9911525}}      

\abstract{This is a set of lectures given at the 99' Cargese Summer School:  Particle Physics : Ideas and Recent
                             Developments.
They contain a pedestrian exposition of recent theoretical progress in
non-perturbative field theory and string theory  based on ideas of duality.}

\begin{document}

\maketitle 

\section{Introduction}

A very important problem in physics is understanding strong 
coupling phenomena.
In the realm of high energy physics an  appropriate example is 
the low energy regime of quantum chromodynamics.
Such examples appear also very frequently in condensed matter systems.

There have been many attempts and methods to attack strong 
coupling problems. These range from qualitative methods, to alternative approximations (non-standard perturbative expansions), to simple truncations
of an exact equation (typically applied to Schwinger-Dyson equations
or renormalization group equations), or finally direct numerical methods
(usually on a lattice).

All methods listed above have their merits, and can be suitable for the appropriate problem. They also have their limitations.
For example , despite the successes of the lattice approach, some questions about QCD still remain today beyond the reach of quantitative approaches. A typical example are dynamic properties like scattering amplitudes.
Consequently, new analytical methods to treat strong coupling problems are always welcome.

The purpose of these lectures was  to communicate to an audience of 
mostly young experimentalists and standard model theorists, the progress in this domain during the past few years.  

The recent understanding of the strongly coupled supersymmetric 
field theories is the starting point of the exposition as well as it central element, electric-magnetic duality.
We will go through the Seiberg-Witten solution for N=2 gauge theories
and we will briefly browse on other developments of these techniques.

The most spectacular impact of these duality ideas has been in string 
theory, a candidate theory for unifying all interactions including gravity.
In string theory, duality has unified the description and scope of distinct string theories.
The importance of new non-perturbative states was realized, and their role in non-perturbative connections was elucidated.
New advances included the first microscopic derivation of the Bekenstein 
entropy formula for black holes.
Moreover, a new link was discovered relating gauge theories to gravity, providing candidates for gauge theory effective strings.
It is fair to say that we have just glimpsed on new structures and connection
in the context of the string description of fundamental interactions.
Whether nature shares this point of view remains to be seen.

There are many excellent reviews that cover some of the topics I present here
and the readers are urgent to complement their reading by referring to them.
I will try to present a short representative list that will be the initial point for those interested to explore the literature.
There are several reviews on supersymmetric field theory dualities \cite{is}-\cite{div}.
Introductory books and lectures in string theory can be found in \cite{GSW}-\cite{pol3}.
Lectures on recent advances and various aspects of non-perturbative string theory 
can be found in \cite{dkl}-\cite{aharony}.

\section{Electric-Magnetic duality in Maxwell theory}

We will describe in this section the simplest realization of a duality symmetry, namely electric-magnetic duality in electrodynamics. We will be employing high energy units $\hbar=c=1$.
The conventional Maxwell equations are 
\be
\vec\nabla\cdot \vec E=\rho \sp \vec\nabla\times\vec B-{\partial \vec E\over \partial t}=\vec J\label{m1}
\ee
\be
\vec \nabla \cdot \vec B=0\sp \vec\nabla\times\vec E+{\partial \vec B\over \partial t}=\vec 0\label{m2}
\ee 

We can use relativistic notation and assemble the electric and magnetic fields into a second rank antisymmetric tensor $F_{\m\n}$ as
\be
E_i=F_{0i}\sp F_{ij}=-\e_{ijk}B^k\sp j^\m=(\rho,\vec J)
\ee
If we define the dual electromagnetic field tensor as 
\be
\tilde F_{\m\n}={1\over 2}\e_{\m\n\rho\s}F^{\r\s}
\ee
Then Maxwell's equations (\ref{m1}),(\ref{m2}) can be written as
\be
\d^{\m}F_{\m\n}=J_{\n}\sp \d^{\m}\tilde F_{\m\n}=0
\ee
The first of these is a true dynamical equation that we will continue to call the Maxwell equation while the second becomes an identity once the fields are written in terms of the electromagnetic potentials, $F_{\m\n}=\d_{\m}A_{\n}-\d_{\n}A_{\m}$.
It is called the Bianchi identity.

Let us first consider the vacuum equations: $\r=0,\vec J=\vec 0$.
They can be written as
\be
\vec\nabla\cdot(\vec E+i\vec B)=0\sp  i\vec\nabla\times(\vec E+i\vec B)+{\partial (\vec E+i\vec B)\over \partial t}=0
\label{compl}\ee
which makes manifest the following symmetry of the equations
\be
\vec E+i\vec B\to e^{i\phi}(\vec E+i\vec B)
\ee
It turns out that only a discrete $\Z_2$ subgroup of this U(1) symmetry ($\phi=\pi/2$) has a chance of  surviving the inclusion of charged matter.
This is known as the electric-magnetic duality transformation 
\be
\vec E\to \vec B\sp \vec B\to -\vec E
\ee
or in tensor form
\be
F_{\m\n}\leftrightarrow \tilde F_{\m\n}
\ee

Once we consider the addition of charges, this symmetry can be maintained only at the expense of introducing also magnetic monopoles. 

The classical (relativistic) equation of motion of a charged particle (with charge $e$)
in the presence of 
an electromagnetic field $F_{\m\n}$ is given by
\be
m~\ddot x^{\m}=e F^{\m\n}\dot x_{\n}
\label{newt}\ee
A magnetic monopole couples to $\tilde F$ in the same way that a charge couples 
to $F$.
Classically, the generalization of the equation above for a particle carrying both an electric charge $e$ and a magnetic charge $g$ is a generalization of (\ref{newt})
\be
 m~\ddot x^{\m}=(eF^{\m\n}+g\tilde F^{\m\n})\dot x_{\n}
\label{newt1}\ee
Classically there are no conceptual changes apart from the fact that the equation of motion is modified.
The reason is that physics classically depends on the field 
strengths rather than gauge potentials.

The situation changes in the quantum theory as was first pointed out by Dirac.
Physics does depend on the potentials rather than field strengths alone, and this provides the famous Dirac quantization condition for the magnetic charge.

An easy way to see this is to write first the classical equation of motion of a
charged particle in the magnetic field $\vec B$ of a magnetic monopole.
\be
m\ddot{\vec  r}=e\dot{\vec r}\times \vec B\sp \vec B={g\over 4\pi}{\vec r\over r^3}
\ee 
We can compute the (semi-classical) rate of change of the orbital angular momentum
\be
{d\vec L\over dt}={d\over dt}\left(m~(\vec r\times \dot{\vec r})\right)=m~\vec r\times \ddot{\vec r}
\ee
Using the equation of motion we can substitute $\ddot {\vec r}$ and find
\be
 {d\vec L\over dt}={eg\over 4\pi}~{\vec r(\dot{\vec r}\times \vec r)\over r^3}=
 {d\over dt}\left({eg\over 4\pi}~{\vec r\over r}\right)
 \ee
This indicates that the conserved angular momentum  is given by
\be
\vec L_{tot}=m(\vec r\times \dot {\vec r})-{eg\over 4\pi}~{\vec r\over r}
\label{ang}\ee
It can be verified that the second piece is the angular momentum of the the electromagnetic field, namely proportional to the spatial integral of the Poynting vector $\vec E\times \vec B$.
Quantization of the total and orbital angular momentum translates via (\ref{ang}) to the Dirac quantization condition
\be
{eg\over 4\pi}=\hbar {n\over 2}\spa\spa\Rightarrow \spa\spa eg =2\pi n\hbar
\label{dirac}\ee
The presence of $\hbar$ in this condition makes obvious that we are 
discussing a quantum effect.
An immediate corrolary is that if a single monopole with charge $g_0$ exists 
then electric charge is quantized in units of  $2\pi\hbar/g_0$.

In general when several electric and magnetic charges are present
the quantization condition reads
\be
e_i~g_j=2\pi \hbar N_{ij}
\ee 
  where $N_{ij}\in \Z$.

\vskip .4cm
\noindent\hrulefill
\vskip .4cm

{\bf Exercise}: Consider a dyon with electric and magnetic charge 
$(e_1,g_1)$ moving in the field of another dyon with charges $(e_2,g_2)$.
Redo the argument with the angular momentum to show that the 
electromagnetic angular momentum is 
\be
\vec L_{em}={(e_2 g_1-e_1 g_2)\over 4\pi}{\vec r\over r}
\ee
which again implies that the 
appropriate
quantization condition here is
\be
e_1 g_2-e_2 g_1=2\pi n ~\hbar
\ee

\vskip .4cm
\noindent\hrulefill
\vskip .4cm

Another point of view is provided by the Dirac string singularity.
As we mentioned above the gauge potential is essential for 
the quantum theory.
$\vec B=\vec \nabla\times \vec A$ implies for a smooth $\vec A$ that 
$\vec \nabla \cdot \vec B=0$.
However, for a point-like magnetic monopole, $\vec \nabla \cdot \vec B\sim \delta^{(3)}(\vec r)$ so that the vector potential must have a 
string singularity.
To put it differently, the existence of a vector potential implies that the magnetic flux emanating from a magnetic monopole must have arrived in some way
at the origin.
This can be done by assuming that we have an infinitely thin solenoid along say the z-axis which brings from infinity the flux emanating from the monopole.
This solenoid which smoothes out the string singularity can be shifted around by gauge transformations. Thus, its position is not a physical observable and one should not be able to measure it.
This was the essence of the original argument of Dirac.
The phase acquired by a charge particle of charge $e$ when transported 
around the solenoid is given by
\be
{\rm phase}=e\oint\vec AS_{\rm solenoid}\cdot d\vec l =eg=2\pi ~{\rm integer}
\ee
which reproduces (\ref{dirac}).  

The upshot of all this is that we can consider including magnetic monopoles in electromagnetism. Then,

$\bullet$ The monopole charge satisfies the Dirac condition.

$\bullet$ The configuration is singular and has an unobservable
 string attached.

\section{Non-abelian gauge theories} 
\setcounter{equation}{0}

The ultraviolet behavior of a U(1) gauge theory is singular (due to the existence of the Landau pole which drives the theory to strong coupling).
It is believed that an IR U(1) gauge theory must be embedded in a spontaneously broken non-abelian gauge symmetry, in order to have regular UV behavior.
 
 We will describe here the fate of Dirac monopoles in the context of the spontaneously broken non-abelian theory.
For the sake  of concreteness we will study the Georgi-Glashow
model.
It is an SU(2) Yang-Mills theory coupled to scalars transforming in the adjoint.
The Lagrangian is
\be
L={1\over 4}F_{\m\n}^aF^{a,\m\n}+{1\over 2}(D_{\m}\phi)^a (D^{\mu}\phi)^a +V(\phi)
\label{act1}\ee   
where
\be
F_{\m\n}^a=\d_{\m}W_{\n}^a-\d_{\n}W^a_{\m}-e~\e^{abc}W_{\m}^bW_{\n}^c
\ee
\be
(D_{\m}\phi)^a=\d_{\m}\phi^a-e~\e^{abc}W^b_{\m}\phi^c
\ee
\be
V(\phi)={\lambda\over 4}(\phi^a\phi^a-a^2)^2
\ee

The minimum of the potential is at $|\phi|^2=\phi^a\phi^a=a^2$.
A vacuum is described  by a solution $\phi^a_0$ of the previous condition.
A solution is characterized by a non-zero three-vector $\phi^a_0$
with length $a$.
This breaks the SU(2) symmetry to U(1). The broken transformations rotate the vacuum vector (Higgs expectation value). The unbroken gauge group corresponds
to rotations that do not change that vector. Obviously this group is composed of the rotations around the vacuum vector and is thus a U(1).

The gauge boson associated to the unbroken U(1) symmetry (that we will call the photon) is $A_{\mu}={\phi_0^a ~W^a_{\mu}\over a}$.
The electric charge (unbroken U(1) generator) is given by
\be
Q={\hbar ~e\over a}\phi_0^a T^a
\ee
where $T^a$ are the $3\times 3$ representation matrices of the adjoint of O(3).

The particle spectrum of this spontaneously broken gauge theory is as follows

\vskip 1cm
\centerline{
\begin{tabular}{|c|c|c|c|}\hline
Particle&mass& spin &electric charge\\ \hline\hline
Higgs&$\sqrt{2\lambda}~a$&0&0\\\hline
$\gamma$&0&1&0\\\hline
$W^{\pm}$&$e~a$&1&$\pm 1$\\\hline
\end{tabular}}
\bigskip

\vskip .4cm
\noindent\hrulefill
\vskip .4cm

{\bf Exercise}: Verify the above.

\vskip .4cm
\noindent\hrulefill
\vskip .4cm

This theory has classical solutions (discovered by 't Hooft \cite{hooft}
and Polyakov \cite{pol}) which are stable and carry magnetic charge under the unbroken U(1).
One has to look for localized solutions to the equation of motion.
Far away the fields must asymptote to those of the vacuum.
In particular the Higgs field $|\phi|\to a$.
We shift the potential so that at the minimum
 the value is zero.
 We can write the Hamiltonian density as
 \be
 H={1\over 2}[\vec E^a\cdot\vec E^a+\vec B^a\cdot\vec B^a+(D_0\phi^a)^2+(D_i\phi^a)^2]+V(\phi)
 \ee
The vacuum is characterized then by $V(\phi)=0$ as well as 
$D_{\mu}\phi^a=0$, $F_{\m\n}^a=0$
 
Such a solution maps the two-sphere at infinity to the Higgs vacuum manifold, which is given by three-dimensional vectors of fixed length.
This is also a two-sphere. The set of smooth maps from $S^2\to S^2$ are classified topologically by their winding number, or their homotopy class and we have $\pi_2(S^2)=\Z$.

The winding number is
\be
w={1\over 4\pi a^3}\int_{S^2}{1\over 2}\e_{ijk}\e^{abc}\phi^a\d_{j}\phi^b\d_k\phi^c ~dS^i
\ee
The magnetic change of the soliton is related to the winding number thus:
\be
g=-{4\pi\over e}w
\label{mono}\ee
This seems not to be the minimal one required by the Dirac 
quantization condition.
One would expect the minimal monopole charge to be $2\pi/e$.
This is explained as follows: we can add fermions in the theory that transform in the spin-1/2 representation (doublet) of SU(2). This would not affect the monopole solution.
On the other hand, now the fermions have U(1) charges that are $\pm e/2$
and they should also satisfy the Dirac condition. This can work only if the minimal magnetic charge is $4\pi/e$ and this is the case.

The solutions with non-trivial winding at infinity must be classically stable since in order to ``unwrap" to a winding zero configuration
they must go through a singularity. Then their kinetic energy becomes infinite, dynamically forbidding their decay.

To find the simplest $w=1$  solution we use the most general spherically symmetric ansatz
\be
\phi^a={x^a\over e~r^2}H(aer)\sp W^a_0=0
\ee
\be
W_i^a=-\e^{aij}{x^j\over er^2}\left[1-K(aer)\right]
\ee
For large $r$, $H\to aer$ while $K\to 0$.
At large distances the configuration for $A_{\mu}$ (the unbroken U(1) gauge field) is exactly the same 
as for a  Dirac monopole.
One would ask: what happened to the Dirac string?
This can be seen as follows: with a singular gauge transformation we can map 
the Higgs field that winds non-trivially at infinity, to one that does not.
Due to the singular gauge transformation the gauge field now acquires a string singularity \cite{hooft}.

\vskip .4cm
\noindent\hrulefill
\vskip .4cm

{\bf Exercise}: Show that in the limit of large Higgs expectation value $a\to \infty$ we recover the Dirac Monopole.

\vskip .4cm
\noindent\hrulefill
\vskip .4cm

We can also construct dyon solutions (as was first done by Julia and Zee \cite{jz}) by allowing $W^a_0$ to be non-zero: 
$W^a_0={x^a\over er^2}J(aer)$.

By manipulating the energy density of a soliton we can derive the following bound for its mass:
\be
M\geq ~a~\sqrt{e^2+g^2}
\ee
where $e$ is the electric charge while $g$ is the magnetic charge.
This bound is known as the Bogomolny'i bound and it is saturated when the potential is vanishing.

In particular, the mass of the monopole in that case is given by
$M=a~g$ and saturates the Bogomolny'i bound. Remembering the Dirac 
quantization condition , $g=4\pi /e$ we obtain
$M=4\pi a/e$   
The mass of the $W^{\pm}$ bosons also saturates the Bogomolny'i bound: $M=a~e$.
In perturbation theory, $e<<1$, the $W$-bosons are much lighter than the monopoles.

Particles and solitons saturating the Bogomolny'i bound are called Bogo\-mo\-lnyi-Prasad-Sommerfield states or BPS states for short.
We have seen that the W-bosons and monopoles are BPS states in the case of zero potential.

The simple model discussed above can be generalized to Yang-Mills
theories with any simple group G coupled to Higgs scalars that break the group to a subgroup H. 
The vacuum again is specified by $V(\phi)=0$, $D_{\mu}\phi=0$. Taking the commutator $[D_{\m}\phi, D_{\n}\phi]=F_{\m\nu}\phi$ we find that the unbroken 
subgroup $H$ is specified from $F_{\m\n}\phi_0=0$.
If the model does not have extra global symmetries or accidental degeneracies
then the vacuum manifold is isomorphic to G/H.
There are non-trivial monopole solutions if $\pi_2(G/H)$ is non-trivial.
From the exact sequence
\be
\pi_2(G)\to \pi_2(G/H)\to \pi_1(H)\to \pi_1(G)\to\pi_1(G/H)\to \pi_0(H)
\ee
one can compute the relevant homotopy group.
We have $\pi_2(G)=0$ for all $G$. When $G$ does not contain U(1) factors $\pi_1(G)=0$ as well so that 
$\pi_2(G/H)=\pi_1(H)$. Thus, there is a winding number (monopole charge) for every unbroken U(1) factor.

\vskip .4cm
\noindent\hrulefill
\vskip .4cm

{\bf Exercise}: Show that the Standard Model does not have smooth monopoles.

\vskip .4cm
\noindent\hrulefill
\vskip .4cm

In the general (G,H) case there is a generalization of the Dirac quantization condition. This has been investigated by Goddard, Nuyts and Olive \cite{gon}
who found that the magnetic charges $g_i$ take values in the weight lattice $\Lambda(H)$  of the unbroken group H.
On the other hand the electric charges $q_i$ take values in the dual of the weight lattice $\Lambda^*(H)$. 
Then the Dirac condition can be written as
\be
e~ \vec q\cdot \vec g= 2\pi N
\ee
with $N\in \Z$.
The dual of the weight lattice is the weight lattice of the dual group $H^{*}$
: $\Lambda^*(H)=\Lambda (H^*)$.
H determines the electric charges while $H^*$ determines the magnetic charges.
Moreover, $(H^*)^*=H$.

For H=SO(3) we have the Dirac quantization condition $e~g=4\pi$.
The dual group $H^*=SU(2)$ with quantization condition $\tilde e~\tilde g=2\pi$.
For SU(N), the dual group is $SU(N)/\Z_N$.

At this point we can describe the $Montonen-Olive$ $conjecture$ \cite{mo}.
A gauge theory is characterized by two groups $H$ and $H^*$. 
There are two equivalent descriptions of the gauge theory.
One where the gauge group is H, the conserved (Noether) currents 
are H-currents, while the $H^*$-currents are topological currents.
In the other the gauge fields belong to the $H^*$
group, the Noether currents are now the topological currents 
of the previous description and vice versa.
Moreover the coupling $q/\hbar$ in the original theory is replaced by $g/\hbar$ in the magnetic theory.
Since $g\sim 1/e$, this conjecture relates a weakly coupled theory to a 
strongly coupled theory.
It is not easy to test this conjecture.
Some arguments were given for this conjecture originally. For example the 
monopole-monopole force was calculated and was dual to the charge-charge force.
However the conjecture cannot be true in a general gauge theory.
In the example of the Georgi-Glashow model the massive charged states 
$W^{\pm}$-bosons have spin 1 and duality maps them to monopoles with spin 0.
One can bypass this difficulty by adding fermions to the model.
Fermions can have zero modes and thus give non-trivial spin to monopoles
making the validity of the conjecture possible.
We need to make monopoles with spin 1. On the way, there will be monopoles also with  spin 0 and 1/2.
This way of thinking leads to N=4 supersymmetric Yang-Mills theory as the prime suspect for the realization of the Montonen-Olive conjecture.

\section{Duality, monopoles and the $\theta$-angle}
\setcounter{equation}{0}

We have seen that for dyons the Dirac quantization conditions reads
\be
q_1 g_2-q_2 g_1=2\pi n
\label{dz}\ee
Let us consider  a pure electric charge $(q,g)=(q_0,0)$ and a generic dyon $(q_m,g_m)$.
Applying (\ref{dz}) we obtain $q_0 g_m=2\pi n$ so that the smallest magnetic charge is $g_{min}={2\pi\over q_0}$. Consider now two dyons with the minimum 
magnetic charge $(q_1,g_{min})$ and $(q_2,g_{min})$.  Applying (\ref{dz}) again we obtain,
\be
q_1-q_2=nq_0
\label{qua}\ee
This is a quantization condition, not for the electric charges but for charge differences.

If we assume that the theory is invariant under CP
\be
(q,g)\to (-q,g)\sp \vec E\to \vec E\sp \vec B\to -\vec B
\ee
then the condition (\ref{qua}) has two possible solutions:
$q=n~q_0$ or $q=\left(n+{1\over 2}\right)q_0$.

Gauge theories have a parameter that breaks CP: the $\theta$-angle.
The addition to the Lagrangian is
\be
L_{\theta}={\theta e^2\over 32\pi^2}\int d^4 x F^a_{\m\n}\tilde F^{a,\m\n}
=\theta~ N
\ee
Where $N\in \Z$ in the integer valued, topological Pontryagin (or instanton) number.
Physics is periodic  in the  $\theta$-angle: $\theta\to \theta+2\pi$ since
$e^{iS'}=e^{iS}e^{2\pi i\theta}=e^{iS}$.

\vskip .4cm
\noindent\hrulefill
\vskip .4cm

{\bf Exercise}: Show that the theory is CP-invariant only for $\theta=0,\pi$.

\vskip .4cm
\noindent\hrulefill
\vskip .4cm

In the presence of the $\theta$-angle there is an ``anomalous" contribution
to the electric charge  of a monopole \cite{witten}
\be
q={\theta e^2\over 8\pi^2}g
\ee 
For a general dyon, one obtains from the Dirac condition
\be
(q,g)=\left(ne+{\theta e\over 2\pi}m~,~{4\pi \over e}m\right)
\label{mn}\ee
where $n,m\in \Z$.
It can be seen that (\ref{mn}) verifies (\ref{qua}).
We can obtain a useful complex representation  by defining 
\be
Q=q+ig=e\left(n+m\left[{\theta\over 2\pi}+i{4\pi\over e^2}\right]\right)
=e(n+m\tau)
\ee
where we have defined the complex coupling constant 
\be
\tau ={\theta\over 2\pi}+i{4\pi\over e^2}
\ee
In this notation the Bogomolny'i bound becomes
\be
M\geq ae ~|n+m\tau|   
\label{bps}\ee

\section{Supersymmetry and BPS states}
\setcounter{equation}{0}

We start with a brief review
of the  representation theory of $N$-extended supersymmetry in 
four dimensions.
A more complete treatment can be found in \cite{BW}.

Supersymmetry is a symmetry that relates fermions to bosons and vice versa. 
Its conserved charges are fermionic (spinors).
For each conserved Weyl spinor charge we have one supersymmetry.
In general we can have more than one supersymmetry 
(extended supersymmetry).

The most general anticommutation relations the supercharges  
can satisfy are \cite{hls}
\be
\{Q_{\a}^I,Q_{\b}^J\}=\e_{\a\b}Z^{IJ}\;\;\;,\;\;\;
\{\bar Q_{\dot\a}^I,\bar Q_{\dot\b}^J\}=\e_{\dot\a\dot\b}\bar Z^{IJ}
\;\;\;,\;\;\;\{Q_{\a}^I,\bar
Q_{\dot\a}^J\}=\delta^{IJ}~2\s^{\mu}_{\a\dot\a}P_{\m}
\,,\label{3999}\ee
where $Z^{IJ}$ is the antisymmetric central charge matrix.
It commutes with all other generators of the super-Poincar\'e
algebra. 

The algebra is invariant under the U(N) $R$-symmetry that rotates
$Q,\bar Q$.
We begin with a description of the representations of the algebra.
We will first assume that the central charges are zero.

$\bullet$ \underline{Massive representations}. We can go to the rest
frame
$P\sim(M,\vec 0)$. The relations become
\be
\{Q_{\a}^I,\bar Q^J_{\dot\a}\}=2M\delta_{\a\dot \a}\delta^{IJ}
\;\;\;,\;\;\;\{Q^I_{\a},Q^J_{\b}\}=\{\bar Q^I_{\dot\a},\bar
Q^J_{\dot\b}\}
=0\,.\label{D1}\ee
Define the 2N fermionic harmonic creation and annihilation operators
\be
A^I_{\a}={1\over \sqrt{2M}}Q^I_{\a}\;\;\;,\;\;\;A^{\dagger
I}_{\a}={1\over \sqrt{2M}}\bar Q^I_{\dot\a}
\,.\label{D2}\ee
Building the representation is now easy. We start with the Clifford
vacuum
$|\Omega\rangle$, which is annihilated by the $A^I_{\a}$ and we
generate
the representation by acting with the creation operators.
There are ${2N}\choose{n}$ states at the $n$-th oscillator level.
The total number of states is $\sum_{n=0}^{2N}$${2N}\choose{n}$, half
of them being bosonic
and half of them fermionic. The spin comes from symmetrization over
the spinorial indices. The maximal spin is the spin of the
ground-states plus $N$.

{\bf Example}. Suppose N=1 and the ground-state transforms into the
$[j]$ representation of SO(3).
Here we have two creation operators.
Then, the content of the massive representation is
$[j]\otimes([1/2]+2[0])
=[j\pm 1/2]+2[j]$.
The two spin-zero states correspond to the ground-state itself and to
the
state with two oscillators.

$\bullet$ \underline{Massless representations}. In this case we can
go to the frame $P\sim (-E,0,0,E)$.
The anticommutation relations now become
\be
\{Q^I_{\a},\bar Q^{J}_{\dot\a}\}=2\left(\matrix{2E&0\cr
0&0\cr}\right)
\delta^{IJ}
\,,\label{D3}\ee
the rest being zero. Since $Q_2^I,\bar Q_{\dot 2}^I$ totally
anticommute,
they are represented by zero in a unitary theory.
We have $N$ non-trivial creation and annihilation operators
$A^I=Q_1^I/2\sqrt{E}$, $A^{\dagger~I}=\bar Q_1^I/2\sqrt{E}$,
and the representation is $2^N$-dimensional.
It is much shorter than the massive one.
Here we will describe some examples (with spin up to one) 
 that will be useful later on.
For N=1 supersymmetry we have the $chiral$ $multiplet$ containing
a complex scalar and a Weyl fermion, as well as the $vector$ $multiplet$
containing a vector and a majorana fermion (gaugino).
In N=2 supersymmetry we have the $vector$ $multiplet$ containing a vector,
a complex scalar and two Majorana fermions, as well as the $hyper-multiplet$,
 containing two complex scalars and two majorana fermions.
Finally in N=4 supersymmetry we have the $vector$ $multiplet$
containing a vector, 4 majorana fermions and six real scalars.

$\bullet$ \underline{Non-zero central charges}. In this case the
representations are massive. The central charge matrix
can be brought by a U(N) transformation to block diagonal form\footnote{We will consider from now on even N.},
\be
\left(\matrix{0&Z_1& 0&0 & &\dots & \cr -Z_1&0&0&0&&\dots&\cr
0&0&0&Z_2&&\dots&\cr 0&0&-Z_2&0&&\dots&\cr
&\dots&\dots&\dots&\dots&\dots&\cr &\dots&&&&0&Z_{N/2}\cr
&\dots&&&&-Z_{N/2}&0\cr}\right)
\,.\label{322}\ee
and we have labeled the real positive eigenvalues by $Z_m$, 
$m=1,2,\ldots,N/2$.
We will split the index $I\to (a,m)$: $a=1,2$ labels positions inside
the $2\times 2$ blocks while $m$ labels the blocks.
Then
\be
\{Q_{\a}^{am},\bar Q_{\dot\a}^{bn}\}=2M\delta^{\a\dot\a}\delta^{ab}\delta^{mn}
\;\;\;,\;\;\;\{Q_{\a}^{am},Q_{\b}^{bn}\}=Z_n\e^{\a\b}\e^{ab}\delta^{mn}
\,.\label{D4}\ee
Define the following fermionic oscillators
\be
A^{m}_{\a}={1\over
\sqrt{2}}[Q^{1m}_{\a}+\e_{\a\b}Q^{2m}_{\b}]\;\;\;,\;\;\;
B^{m}_{\a}={1\over \sqrt{2}}[Q^{1m}_{\a}-\e_{\a\b}Q^{2m}_{\b}]
\,,\label{D5}\ee
and similarly for the conjugate operators.
The anticommutators become
\be
\{A^m_{\a},A^n_{\b}\}=\{A^m_{\a},B^n_{\b}\}=\{B^m_{\a},B^n_{\b}\}=0
\,,\label{D6}\ee
\be
\{A^{m}_{\a},A^{\dagger
n}_{\b}\}=\delta_{\a\b}\delta^{mn}(2M+Z_n)\;\;\;,\;\;\;
\{B^{m}_{\a},B^{\dagger n}_{\b}\}=\delta_{\a\b}\delta^{mn}(2M-Z_n)
\,.\label{D7}\ee
Unitarity requires that the right-hand sides in (\ref{D7}) be
non-negative.
This in turn implies the bound
\be
M\geq {\rm max}\left[ {Z_{n}\over 2}\right]
\,.\label{D8}\ee
which turns out to be no other than the Bogomolny'i bound.
Supersymmetry in this sense ``explains" the Bogomolny'i bound:
it is essential for the unitarity of the underlying theory.

Consider $0\leq r\leq N/2$ of the $Z_n$'s to be equal to $2M$.
Then $2r$ of the $B$-oscillators vanish identically and we are left
with $2N-2r$ creation and annihilation operators.
The representation has $2^{2N-2r}$ states.
The maximal case $r=N/2$ gives rise to the short BPS multiplet whose
number of states are the same as in the massless multiplet.
The other multiplets with $0<r<N/2$ are known as intermediate BPS
multiplets.

BPS states are important probes of non-perturbative physics in
theories with extended ($N\geq 2$) supersymmetry.
The BPS states are special for the following reasons:

$\bullet$ Due to their relation with central charges, and although
they are massive, they form multiplets under extended SUSY which are
shorter than the generic massive multiplet.
Their mass is given in terms of their charges and Higgs (moduli) expectation
values.

$\bullet$ They are the only states that can become massless when we vary coupling constants and Higgs expectation values.

$\bullet$ When they are at rest they exert no force on each other.

$\bullet$ Their mass-formula is supposed to be exact if one uses
renormalized values for the charges and moduli.\footnote{In
theories with $N\geq 4$ supersymmetry there are no
renormalizations.}
The argument is that quantum corrections would spoil the relation of
mass and charges, and if we assume unbroken SUSY at the quantum level
there would be incompatibilities  with the dimension of their
representations.

$\bullet$ At generic points in moduli space (space of couplings and Higgs expectation values)  they are stable.
The reason is the dependence of their mass on conserved charges.
Charge and energy conservation prohibits their decay.
Consider as an example, the BPS mass formula
\be
M^2_{m,n}={|m+n\tau|^2\over \tau_2}\;\;,
\label{du2}
\ee
where $m,n$ are integer-valued conserved charges, and $\tau$ is a
complex modulus. We have derived this  BPS formula 
in the context of the SU(2) gauge theory.
Consider a BPS state with charges $(m_0,n_0)$, at rest, decaying into N
states
with charges $(m_i,n_i)$ and masses $M_i$, $i=1,2,\cdots,N$.
Charge conservation implies that $m_0=\sum_{i=1}^N m_i$,
$n_0=\sum_{i=1}^N n_i$.
The four-momenta of the  produced particles are $(\sqrt{M_i^2+\vec
p_i^2},\vec p_i)$ with $\sum_{i=1}^N \vec p_i=\vec 0$.
Conservation of energy implies
\be
M_{m_0,n_0}=\sum_{i=1}^N\sqrt{M_i^2+\vec p^2_i}\geq \sum_{i=1}^N
M_i\;\;.
\label{du1}\ee
Also in a given charge sector (m,n) the BPS bound implies that any mass
$M\geq M_{m,n}$, with $M_{m,n}$ given in (\ref{du2}).
From (\ref{du1}) we obtain
\be
M_{m_0,n_0}\geq \sum_{i=1}^N M_{m_i,n_i}\;\;,
\label{du3}
\ee
and the equality will hold if all  particles are BPS and are produced
at rest ($\vec p_i=\vec 0$).
Consider now the two-dimensional vectors $v_i=m_i+\tau n_i$ on the
complex $\tau$-plane, with length $||v_i||^2=|m_i+n_i\tau|^2$.
They satisfy $v_0=\sum_{i=1}^N v_i$.
Repeated application of the triangle inequality implies
\be
||v_0||\leq \sum_{i=1}^N ||v_i||\;\;.
\label{du4}
\ee
This is incompatible with energy conservation (\ref{du3}) unless
all vectors $v_i$ are parallel. This will happen only if $\tau$ is
real which means when $e=\infty$ a highly degenerate case.
For energy conservation it should also be a rational number.
Consequently, for $\tau_2$ finite, the BPS
states of this theory are absolutely stable. This is always true in
theories
with more than N$>2$
supersymmetry in four dimensions.
In cases corresponding to theories with 8 supercharges, there are
regions in the moduli space, where BPS states, stable at weak coupling,
can decay at strong coupling. However, there is always a large region
around weak coupling where they are stable.

\section{Duality in N=4 super Yang-Mills theory}
\setcounter{equation}{0}

The four-dimensional quantum field theory with maximal supersymmetry 
is the N=4 Yang-Mills theory.\footnote{More than four supersymmetries in four dimensions imply the existence of spins bigger than one and thus  non-renormalizability.
Such theories are good as effective field theories.} 
The action of N=4 Yang-Mills is completely specified by the choice of the gauge group G (that we will assume simple here).
As pointed out in a previous section, the only N=4 multiplet with spin at most one is the vector multiplet.
The particle content is a vector multiplet in the adjoint of the gauge group containing a vector, four fermions and six scalars.
There is an $SU(4)\sim O(6)$ global symmetry (the R-symmetry). The supercharges transform in the {\bf 4}, as well as the fermions, while the scalars transform in the {\bf 6} (vector of O(6)).
The kinetic terms of various particles as well as their couplings to the 
gauge field are standard.
The Lagrangian is 
\def\Dslash{{D\hspace{-.22cm}/}}
\be
L_{N=4}=-{1\over 4g^2}~Tr~\left[F_{\m\n}F^{\m\n}+\bar \chi^i\Dslash \chi^i
+D_{\m}\phi_a D^{\m}\phi_a+{\rm Yukawa~~terms}
+\right.
\ee
$$\left.+[\phi_a,\phi_b^{\dagger}] [\phi_a,\phi_b^{\dagger}]\right]
+
{\theta\over 32\pi^2}Tr~F\tilde F
$$
The minima of the scalar potential are given by $[\phi_a,\phi_b^{\dagger}]=0
$ and they are solved by a scalar belonging in the Cartan(G).

\vskip .4cm
\noindent\hrulefill
\vskip .4cm

{\bf Exercise}: Show that for a generic Higgs expectation value in the Cartan of G, the gauge group G is broken to the abelian Cartan{G}.

\vskip .4cm
\noindent\hrulefill
\vskip .4cm

This is the generic Coulomb phase where the massless gauge bosons are $N_c$
photons, where $N_c$ is the rank of G.
The massive W-bosons are electrically charged under the Cartan(G).
Their masses saturate the BPS bound and they are 1/2-BPS states (the shortest representations, as  short as the massless).
There are also 1/2-BPS 't Hooft -Polyakov monopoles in the theory.

The N=4 1/2-BPS mass formula is
\be
M^2={1\over \tau_2}|\vec \phi\cdot (\vec n+\tau\vec m)|^2
\ee
with $\tau={\theta\over 2\pi}+i{4\pi\over g^2}$.
$\vec\phi$ is the vev of the Higgs, while $\vec n,\vec m$ are the integers specifying the electric and magnetic charges respectively.

We will further set $G=SU(2)$ for simplicity. The generalization to other groups is straightforward.

N=4 super yang-Mills for any gauge group is a scale invariant theory 
\cite{n=4}. Its 
$\beta$-function is zero non-perturbatively.
Moreover its low energy two-derivative effective action has no quantum corrections (even beyond perturbation theory). This does not imply, however ,
that the theory is trivial. Correlation functions are non-trivial and it is an open problem to compute them exactly (apart from some three point functions
protected by non-renormalization theorems).

Here the monopoles are in BPS multiplets similar to those of the W-bosons and 
the Montonen-Olive duality has a chance of being correct.
For $\theta=0$ it involves inversion of the coupling constant $g\to {4\pi\over g}$ as well as interchanging of electric and magnetic charges
$n\to m,m\to -n$.
If this is combined with the periodicity in $\theta$: $\theta\to \theta+2\pi$
we obtain an infinite discrete group, SL(2,$\Z)$.
It can be represented by $2\times 2$ matrices with integer entries and unit determinant
\be
\left(\matrix{a&b\cr c&d}\right)\sp ad-bc=1\sp a,b,c,d \in \Z
\ee
The associated transformations act as
\be
\tau\to {a\tau+b\over c\tau +d}\sp \left(\matrix{n\cr m}\right)=
 \left(\matrix{a&b\cr c&d}\right)\left(\matrix{n\cr m}\right)
 \ee
 There are two generating transformations: $\tau\to \tau+1$ (periodicity in $\theta$) and strong-weak coupling duality $\tau\to -1/\tau$.
 
 \vskip .4cm
\noindent\hrulefill
\vskip .4cm

{\bf Exercise}: Show that the BPS mass formula is invariant under the SL(2,$\Z$)
duality.

\vskip .4cm
\noindent\hrulefill
\vskip .4cm

Can we test Montonen-Olive duality? There are some further indications
that it is valid:

$\bullet$ In perturbation theory we have states
with electric charge $\pm 1$ (the W-boson multiplets).
Then SL(2,$\Z$) duality predicts the existence of dyons with charges
\be
\left(\matrix{a&b\cr c&d}\right)\left(\matrix{1\cr 0}\right)=
\left(\matrix{a\cr c}\right)
\ee
where the greatest common divisor of a,c is one, $(a,c)=1$.
All such dyons must exist, if M-O duality is correct.
For example, we have seen that the (0,1) state, the magnetic monopole, exists in the non-perturbative spectrum.
On the other hand no (0,2) monopole should exist, but the dyon (1,2) should exist.
This is a subtle exercise in geometry and quantum mechanics: one has to show that an appropriate supersymmetric quantum mechanical system on a non-trivial quaternionic manifold (the moduli space of dyons with a given magnetic charge) has a certain number of normalizable ground states.
This in turn transforms into the question of existence of certain forms in the moduli space.
This test has been performed successfully for magnetic charge two \cite{sen} 
and the general case in \cite{segal}.

$\bullet$ There is a relatively simple object to compute in a supersymmetric quantum field theory, namely the Witten index.
This amounts to doing the path integral on the torus with periodic boundary conditions for bosons and fermions.
On such a flat manifold the result is a pure number that counts the supersymmetric ground states. 
If however, the path integral is performed on a non trivial 
compact or non-compact manifold with supersymmetry preserving boundary conditions, then the Witten index depends non-trivially both on the manifold and the coupling constant $\tau$.
The Witten index for N=4 Yang-Mills was computed \cite{vawi} on K3 and on ALE manifolds and gave a result that was covariant under SL(2,$\Z$) duality.

$\bullet$ In string theory, the M-O duality of N=4 super-Yang Mills is equivalent to T-duality (a perturbative duality of string theory that is well understood) via a string-string duality that has had its own 
consistency checks.

At this point we should consider the question whether it makes sense to expect that we can have a way to prove something like M-O duality.
In order for this question to be meaningful, there must be an alternative way of defining the non-perturbative (strongly coupled theory).
Duality can be viewed as a different (independent) definition of the strong coupling limit and in that case it makes sense to ask whether the two non-perturbative definitions agree.
Unfortunately for supersymmetric theories we do not have a non-perturbative definition. The obvious and only such definition (lattice) breaks supersymmetry and remains to be seen if it can be used in that vein.

Montonen-Olive duality can be viewed as a (motivated and possibly incomplete) definition of the non-perturbative theory.
As with any definition it must satisfy some consistency checks.
For example if a quantity satisfies a non-renormalization theorem and can be thus computed in perturbation theory, it should transform appropriately under duality, etc.
In all cases of duality in supersymmetric field and string theories we are checking their consistency rather than proving them.

\section{N=2 supersymmetric gauge theory}
\setcounter{equation}{0}

The two relevant N=2 massless multiplets are the vector multiplet and the hypermultiplet.
Here we will consider the simplest case: pure gauge theory, with vector multiplets only. Hypermultiplets can also be accommodated but we will not 
discuss them further here.
The vector multiplet $(A_{\m}^a,[\chi^a,\psi^a],A^a)$ contains a vector, two majorana spinors and a complex scalar $A^a$ all in the adjoint of the gauge group.

The renormalizable N=2 Lagrangian is 
\be
L_{N=2}={1\over g^2}Tr\left[-{1\over 4}F_{\m\n}F^{\m\n}+(D_{\mu}A)^{\dagger}D^{\m}A-{1\over 2}[A,A^{\dagger}]^2
-i\psi\s^{\mu}D_{\m}\bar \psi-
\right.\ee
$$
\left.-i\chi\bar \s^{\mu}D_{\m}\bar \chi-i\sqrt{2}[\psi,\chi]A^{\dagger}
-i\sqrt{2}[\bar\psi,\bar\chi]A\right]+{\theta\over 32\pi^2}Tr F_{\m\n}\tilde F^{\m\n}
$$
This defines the ultraviolet theory. The theory is asymptotically free 
and it flows to strong coupling in the infrared.
The minima of the potential are as before: $A$ must take values 
in the Cartan of the gauge group.
The values at the Cartan are arbitrary (flat potential) and are moduli of the problem. Put otherwise, there is a continuum of vacua specified by the expectation values of the Higgs in the Cartan.
A non-zero (generic) Higgs expectation value breaks the gauge group to the Cartan, $U(1)^{N_c}$ and we are in the Coulomb phase.
The $G/U(1)^{N_c}$ vector multiplets become massive (W-multiplets) and are BPS multiplets of N=2 supersymmetry since they have the same number of states as the massless multiplets.
There are monopoles as usual since $\pi_2(G/U(1)^{N_c})=\Z^{N_c}$.

From now on we specialize to G=SU(2) to avoid unnecessary complications.
Other groups can be treated as well.

The fundamental question we would like to pose here concerns strong coupling.
We have mentioned that the theory is asymptotically free.
If one is interested in physics at low energy then he has to 
solve a strong coupling problem.
As we will see, supersymmetry here will help us to solve this problem.
The end result will be the exact two-derivative Wilsonian effective action at low energy.
Obviously, the low energy effective action is something easy to calculate in an IR-free theory since one can use perturbation theory (e.g. QED).

The Wilsonian effective action at a scale $E_0$ is constructed by integrating out degrees of freedom with energy $E\geq E_0$.

Going a bit back we can ask: what is the low energy effective 
action for the N=4
super Yang-Mills discussed in the previous section, in the Coulomb phase.
We have seen that the W-bosons are massive with masses $\sim |\phi_0|^2$.
If we are interested in energies smaller than their mass we can integrate them out.
The low energy theory will contain only the photon multiplets with possible 
extra interactions induced by the massive particles in the loops.
It turns out, however, that N=4 supersymmetry protects the two-derivative effective action from corrections due to quantum effects (even beyond perturbation theory).
The most important part in the IR, the two-derivative action, again describes free photon multiplets with no additional interactions.
Moreover it is known that the four-derivative terms (like $F^4$ terms) obtain corrections only from one loop in perturbation theory (in four dimensions).

We would like to solve the same problem in the  N=2 gauge theory, where
the two-derivative effective action does get quantum corrections from massive states.
In this theory, the W-multiplets are massive with BPS masses $m^2=|A|^2$ where 
$A$ is the third component of the non-abelian scalar which parameterizes the moduli space (a copy of the complex plane).
We would like to integrate out the W-bosons and find the effective physics for the photon multiplet for energies well below the W mass $|A|$.
The effective action will of course be of the non-renormalizable type,
a fact  acceptable for an effective theory.
The low energy effective action will contain a photon, two photinos and a complex scalar $A$.

There are two special points in the space of vacua (moduli space).

$\bullet$ $A=0$. Here the gauge symmetry is enhanced to SU(2), since the W-bosons become massless.

$\bullet$ $A\to \infty$. This is the abelian limit and as we will see we can trust perturbation theory in that neighborhood of moduli space.

An important point to make  is that we do not expect the N=2 supersymmetry to break. Consequently, the effective field theory could be one of the most  general N=2 theories with a single vector multiplet.
The most general such (non-renormalizable) action is known.
It depends on a single unknown function ${\cal F}$ known as the prepotential 
which is a holomorphic function of the complex scalar $A$.
We summarize it below.
\be
L_{eff}\sim Im~{\partial ^2\f\over \partial A^2}\left[-{1\over 4}F_{\m\n}F^{\m\n}+D_{\m}A D^{\m}A^{\dagger}\right]+Re~{\partial ^2\f\over \partial A^2}{1\over 32\pi}F_{\m\n}\tilde F^{\m\n}+{\rm fermions}
\ee
As obvious from above $ Im~{\partial ^2\f\over \partial A^2}$ is the inverse effective coupling while $Re~{\partial ^2\f\over \partial A^2}$ is 
the effective $\theta$-angle.
It is obvious that if we manage to find $\f(A)$ we have completely determined the low-energy effective action.

Classically (at the tree level) $\f(A)={1\over 2}\tau A^2$ reproduces the 
classical (UV) coupling constant $\tau$.
The prepotential $\f(A)$ will have both perturbative and non-perturbative corrections
(coming here from instantons).

An important ingredient of the effective U(1) theory is the value of the central charge (that determines the BPS formula) as a function of the modulus $A$:
\be
Z=A~n_e+ {\partial \f\over \partial A}~n_m
\label{z}\ee
where $n_e,n_m$ are integers that determine the electric and magnetic charges 
respectively.
Here we see an example where the central charge receives quantum corrections
(since $\f$ does) but the mass equality $M=|Z|$ for BPS states still remains valid. This happens because the mass is also renormalized as to keep the
BPS  relation valid.

At tree level we have
\be
Z_{\rm tree}=A(n_e+\tau n_m)
\ee
We will define the dual Higgs expectation value $A_D\equiv {\partial \f\over \partial A}$.
Then we have the following M-O-like SL(2,$\Z$) duality:$A\leftrightarrow A_D$,
$n_e\leftrightarrow n_m$.

We need a better coordinate than A on the moduli space.
The reason is that A is not gauge invariant. The Weyl element of the original SU(2) gauge group acts as $A\to -A$.
Thus, a gauge-invariant coordinate is $u=A^2/2$. At $A=u=0$ we have gauge symmetry enhancement $U(1)\to SU(2)$.

\subsection{The fate of global symmetries}

An N=2 supersymmetric theory has a $U(2)=U(1)\times SU(2)$ (global) 
R-symmetry that rotates the two supercharges.
The various fields of the vector multiplet transform as follows:

\vskip 1cm
\centerline{
\begin{tabular}{|c|c|c|}\hline
Particle&U(1)&SU(2)\\ \hline\hline
$A_{\mu}$&0&singlet\\\hline
$\chi,\psi$&1&doublet\\\hline
A&2&singlet\\\hline
\end{tabular}}
\bigskip

The U(1) R-symmetry has a chiral anomaly, which means that it is broken by instanton effects.
For a gauge group SU(N)\footnote{We will do this analysis for general N
although eventually we will be interested in N=2.}
 an instanton has a zero mode for each left
fermion in the fundamental and 2N zero modes for a fermion in the adjoint.
Here our fermions are in the adjoint.
In order to obtain a non-zero amplitude in an instanton background we need to soak the fermionic zero modes, and that can be done by inserting the appropriate number of fermion operators in the path integral.
We thus obtain that the simplest non-vanishing correlator is
\be
G=\langle \prod_{i=1}^{2N}\chi(i)\prod_{i=1}^{2N}\psi(i)\rangle \not =0
\ee
$G$ has U(1) charge 4N and transforms under a U(1) transformation $e^{ia}$ as $G\to e^{i4Na}G$. This implies that since $G\not=0$, the U(1) symmetry is broken to $\Z_{4N}$.
The unbroken global symmetry is $SU(2)\times \Z_{4N}$.
However, the center of SU(2) (that acts as $(\psi,\chi)\to -(\psi,\chi)$)
is contained in $\Z_{4N}$.
We conclude that the global symmetry is   $(SU(2)\times \Z_{4N})/\Z_2$.
When we have a non-zero Higgs expectation value $A$, the global symmetry breaks further.
For example in the SU(2) case $u\sim A^2$ has charge 4 under $\Z_{8}$ so that $\Z_{8}$ breaks to $\Z_{4}$.
For SU(2) this is the end of the story and the unbroken global symmetry is 
 $(SU(2)\times \Z_{4})/\Z_2$.
 The broken $\Z_8$ acts as $u\to -u$.

\vskip .4cm
\noindent\hrulefill
\vskip .4cm

{\bf Exercise}: Find the unbroken global symmetry for G=SU(3), SU(4).

\vskip .4cm
\noindent\hrulefill
\vskip .4cm

\subsection{The computational strategy}

We need to calculate the holomorphic prepotential $\f(u)$ in order to 
determine the exact effective action.
The central idea is that if we know the singularities and monodromies of a
holomorphic function then there is a concrete procedure that reconstructs it.

The strategy is \cite{sw} to find the singularities and monodromies of $\f(u)$.

$\bullet$ Use perturbation theory to study the singularity at $u\to \infty$.

$\bullet$ Use physical arguments and local SL(2,$\Z$) duality  to determine the 
behavior at the other singular points.

$\bullet$ Use math techniques to reconstruct $\f(u)$.

Classically the only two singular points are $A\to \infty$ and $A\to 0$ where we have gauge symmetry enhancement and the U(1) effective theory breaks down.

\subsection{Perturbation Theory}

An important ingredient in perturbation theory is that the two-derivative 
effective action obtains corrections only at one loop (in the presence of unbroken N=2 supersymmetry) \cite{n=2}.

There is a simple but ``dirty" argument. The (anomalous) divergence of the R-current $\partial_{\m}J_R^{\mu}\sim F\tilde F$ belongs to the same N=2 supermultiplet 
with the trace of the energy-momentum tensor $T_{\m\n}$.
Classically  the theory is scale invariant and $T_{\m\n}$ is traceless.
, quantum effects break scale invariance and in the quantum theory the trace is proportional to the $\beta$-function of the theory.
On the other hand the axial anomaly obeys an Adler-Bardeen non-renormalization theorem that specifies that in a given scheme (the Adler-Bardeen scheme)
it receives quantum corrections at one loop only.
Unbroken N=2 supersymmetry implies that this is also true for the 
$\beta$-function of the theory and consequently for the prepotential.
The dirtyness of the argument has to do with subtleties about renormalization.
If that were not the case a similar argument would work for N=1 gauge theories.
It is known however that, generically, in N=1 gauge theories with matter, the
beta function obtains corrections at all loops.
The reason is that in N=1 theories the Adler-Bardeen current and the one that belongs to the same multiplet as the trace of the energy-momentum tensor are different due to renormalization.
This argument works though for N=2 theories, \cite{n=2}.

Thus, we are left with a one-loop calculation to do.

The one-loop $\beta$- function in field theory is given by the following formula
\be
\mu{\partial\over \partial \m}g^{\rm eff}(\mu)\equiv \beta(g)
\ee
\be
{1\over g^2(\m)}={1\over g_0^2}-{1\over 8\pi^2}\sum_i b_i~\log\left({\mu^2+m_i^2\over \Lambda^2}\right)
\label{beta}\ee
where the $\beta$-function coefficients are given by 
\be
b_i=(-1)^{2s} Q^2\left({1\over 12}-s^2\right)
\ee
Here $s$ is the helicity and $Q$ is an appropriately normalized generator of the gauge group.
A boson contributes 1/12, a Weyl fermion 1/6 while a vector contributes -11/12.
The summation is over all particles, with masses $m_i$.
Expression (\ref{beta}) is approximate at the thresholds (when $\mu$ comes near to one of the masses $m_i$) but very accurate elsewhere.

Assume for simplicity that there is only one particle with mass m contributing to the $\beta$-function.
The following behavior of the effective coupling can be seen from (\ref{beta}):

$\bullet$ For $\mu >>m$ there is logarithmic running.

\begin{figure}
\begin{center}
\vspace{0cm}  
\epsfig{file=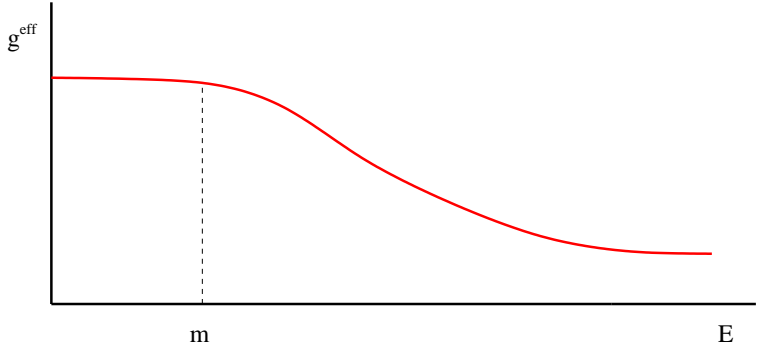,width=10cm}
\caption{The running coupling past a threshold. }
\label{fig1}
\end{center}
\end{figure}

$\bullet$ For $\mu<<m$ the coupling ``freezes" at the value $g^{-2}(m)=g^{-2}_0-{b\over 8\pi^2}\log{m^2\over \Lambda^2}$.
This is reasonable since for energies lower than $m$ all contributions of the particle have been integrated out. Consequently there is no further running of the coupling. This behavior is portrayed in Fig. \ref{fig1}.

The massive particles we are integrating out are two massive vector multiplets.
Their mass is $m=|A|$.
The contribution of a single vector multiplet to the $\beta$-function coefficient is $b_{v}=2{1\over 12}+4{1\over 6}-2{11`\over 12}=-1$.
The electric charge is 1 so that in total $b=-2~Q^2=-2$.
Since we integrate out all energies above the mass of the particles the effective coupling for energy below $|A|$ is frozen to 
\be
{1\over g_{\rm eff}^2}={1\over g_0^2}+{2\over 8\pi^2}\log{|A|^2\over \Lambda^2}
\label{beta2}\ee
We can absorb $g_0$ into $\Lambda$ (dimensional transmutation)  and rewrite
\be
{1\over g_{\rm eff}^2}={1\over 4\pi^2}\log{|A|^2\over \Lambda^2}
\label{beta3}\ee
This must come from a holomorphic prepotential $\f(A)$ so that 
\be
{Im\f''(A)\over 4\pi}={1\over 4\pi^2}\log{|A|^2\over \Lambda^2}
\ee
The solution is
\be
\f(A)={i\over 2\pi}A^2\log{A^2\over \Lambda^2}
\label{prep1}\ee
By allowing $\Lambda$ to be complex, we can absorb into it the classical
$\theta$-angle.
t one loop
\be
\left.\theta_{\rm eff}\right|_{\rm one-loop}=4(Arg(\Lambda)-Arg(A))
\ee

In what region of the moduli space can we trust perturbation theory?
This can be seen from Fig. 1. Now $m=|A|$. By taking $|A|$ larger and larger
while keeping $\Lambda$ (the UV coupling) fixed, 
the effective coupling freezes at lower and lower values.
Thus, in the neighborhood of $A=\infty$ perturbation theory is reliable.

As can be seen from the one-loop prepotential there are two singularities:
$A=0$ and $A=\infty$.
The singularity at $A=\infty$ we trust since perturbation theory is a good guide there. This is not true for the one at $A=0$ where the theory is strongly coupled.
Can this be the only singularities of the prepotential?
The answer is no, for the following reasons:
A holomorphic function with two singularities on the complex plane, 
and a logarithmic cut at $\infty$ (remember that we trust this) is unique and given by the one-loop result. 

On the other hand, this is incompatible for two reasons.

$\bullet$ For smaller values of $A$, the coupling constant becomes negative.

$\bullet$ The one-instanton contribution to the $\beta$-function had been computed before and found to be non-zero.

The only way out is to assume that $\f(A)$ has more singularities 
on the complex plane.

\subsection{Singularities and monodromy}

\begin{figure}
\begin{center}
\vspace{0cm}  
\epsfig{file=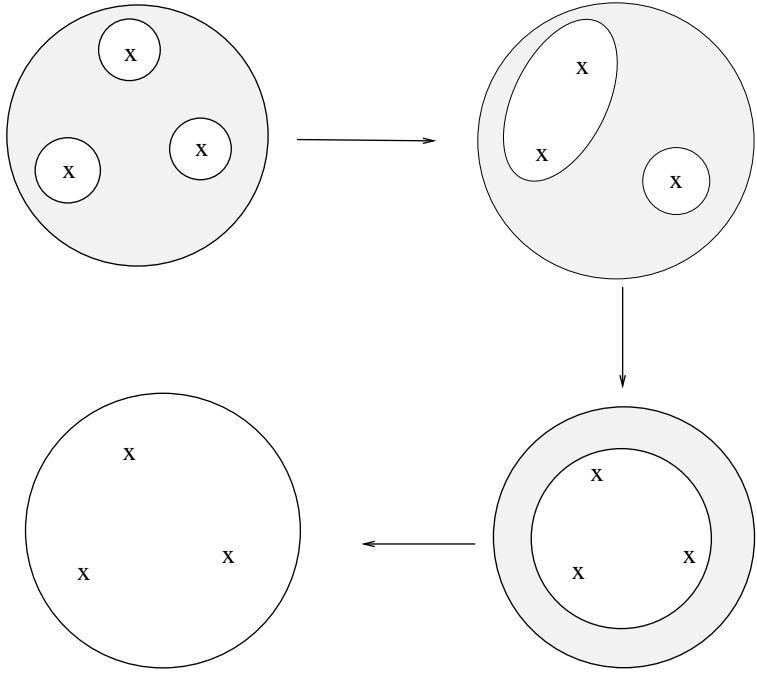,width=10cm}
\caption{The global monodromy condition}
\label{fig2}
\end{center}
\end{figure}

Consider the complex function $f(z)=\sqrt{z}$. If we encircle once the 
origin, $z\to e^{2\pi i}~z$, then $f(e^{2\pi i}z)=-f(z)$.
Thus the function does not return back to itself. This is a signal that the point $z=0$ is a singular point for the function, in this case the start of a branch cut.
The behavior of a complex function or a set of functions after transport around a point (singularity) is called the monodromy.
In general a set of functions, transported once around the singular  
point $z_0$ return to
a linear combination of themselves. We write
\be
F_i((z-z_0)e^{2\pi i})=M_{ij}(z_0)F_{j}(z)
\label{monodr}\ee
 The matrix $M$ depends on the singular point, and is called the monodromy matrix at that point.
 Monodromy has a topological character. The monodromy matrices do not change
 under smooth deformations of the contour. Non-smooth deformations include the contour crossing another singular point. 
 
This matrix is important because it plays an essential role in the Riemann-Hilbert problem: if we know the position of the singularities and the monodromy around each one, of a set of holomorphic functions, then we can reconstruct them uniquely.

If we want to be a bit more careful then we will realize that $\f(a)$ is not really a function. We have seen earlier that SL(2,$\Z$) duality interchanges
the derivative of $\f$, $A_D$ with $A$.
The relevant holomorphic objects to consider are the pair $A$ and $A_D$ viewed 
both as functions of the good coordinate $u=A^2/2$.
If we make a circle around $u=0$, then $u\to e^{2\pi i}u$ and $A\to -A$.
\be
A_D=\f'(A)={2iA\over \pi}\left(\log{A\over \Lambda}+{1\over 2}\right)
\ee
Thus, when $A\to -A$ then
\be
A_D\to \f'(-A)= -{2iA\over \pi}\left(\log{-A\over \Lambda}+{1\over 2}\right)
=-A_D+2A
\ee
Thus, the monodromy around $u=0$ is given by
\be
\left(\matrix{A_D\cr A}\right)\to M_{0}\left(\matrix{A_D\cr A}\right)=\left(\matrix{-1&2\cr0&-1}\right)\left(\matrix{A_D\cr A}\right)
\ee
Similarly, the monodromy around $u=\infty$ is
\be
\left(\matrix{A_D\cr A}\right)\to M_{0}\left(\matrix{A_D\cr A}\right)=\left(\matrix{-1&2\cr 0&-1}\right)\left(\matrix{A_D\cr A}\right)
\ee
 The two matrices satisfy $M_0~M_{\infty}=1$.
 This is a general property of monodromy. If we have a number of singularities on the sphere then the associated monodromy matrices satisfy $\prod_i~ M_i=1$.
 The proof of this is sketched in Fig. 2.
 We start with a number of independent contours that we can deform until we obtain a single one that we can shrink to zero on the back side of the sphere.

As we mentioned above, if we only have two singularities then the perturbative result is the whole story. We had argued though that instanton corrections are non-trivial.
We will analyze now their expected form.
From the one-loop running we have 
$$g^2(A)={4\pi^2\over \log{A^2\over \Lambda^2}}$$
The k-th instanton contribution is proportional to $exp[-k{8\pi^2\over g^2}]
\sim \left({\Lambda \over A}\right)^{4k}$.
This breaks the U(1) R-symmetry as expected (A is charged).
We can restore the U(1) symmetry if we allow $\Lambda$ to transform with charge 2.
Then the exact prepotential is expected to have the following form,
\be
\f(A)={i\over 2\pi}\log{A^2\over \Lambda^2}
+A^2\sum_{k=1}^{\infty}~c_k\left({\Lambda \over A}\right)^{4k}
\ee
One needs to calculate the coefficients $c_k$.

We have seen that we need more singularities than the ones we have observed in perturbation theory.
The possible meaning of such singularities would be that they are due to states that become massless at that particular point of the moduli space.
This would signal the breakdown of the effective theory, since we have integrated out something very light. 
There are two possibilities; the particles that become massless are in vector multiplets or in hypermultiplets.
The guess of Seiberg and Witten is that only the second case is correct. First we have an abundance of non-perturbative hypermultiplets, namely monopoles and dyons that could in principle become massless at strong coupling.
There are various arguments that indicate that it is implausible that vectors become massless \cite{sw}.

One extra constraint is that singularities that appear on the sphere except the points $A=0$ and $A=\infty$ must appear in pairs. The reason is that if a singularity appear at $u=u_0$ then by the broken R-symmetry it must be that also $u=-u_0$ is a singularity.
The minimal number of singularities we need is three. Since $A=\infty$ is a singularity, we must also have a pair of singularities in the interior of the moduli space.
In that case, the classical singularity at $A=0$ must be absent non-perturbatively.
These assumptions can be verified a posteriori.

We put two extra singularities, one at $u=\Lambda^2$ and another at $u=-\Lambda^2$ (this can be thought of as a non-perturbative definition of 
$\Lambda$.
A natural guess for the  particle that becomes massless at $u=\Lambda^2$
is that it is the monopole. However there are monodromy 
constraints that must be satisfied and we must take them into account.

We will assume that some dyon becomes massless at a given point of the moduli space and try to compute the monodromy matrix.
The low energy theory around the singularity must include the very light dyon.
Then we would like to compute the local coupling by computing a one-loop diagram where the dyon is going around the loop.
This is not obvious how to do.
It is duality at that point that comes to the rescue.

\subsection{The duality map}

We will need the following identities in four dimensions
\be
F_{\m\n}F^{\m\n}=-\tilde F_{\m\n}\tilde F^{\m\n}\sp \tilde{\tilde F}=-F
\ee
The quadratic action can be written as
\be
S= {1\over 32\pi}Im \int \tau(a)(F+i\tilde F)^2={1\over 32\pi}Im \int \tau(a)(2F^2+2iF\tilde F)
\ee
If we want to consider $F$ as an independent variable we must explicitly impose the Bianchi identity $dF=0$.
This we can do by adding an extra term in the action
\be
\Delta S={1\over 8\pi}\int V_{\m}\e^{\m\n\r\s}\partial_{\n}F_{\r\s}
\ee
Integrating over the vector $V_{\mu}$ gives a $\delta$-function that 
imposes the Bianchi identity.
$\Delta S$ can be rearranged as follows
\be
\Delta S=-{1\over 8\pi}\int \partial_{\m}V_{\nu} \e^{\m\n\r\s}F_{\r\s}
=-{1\over 8\pi}\int F\tilde F_D={1\over 16\pi}Re\int (\tilde F_D-i F_D)(F+i\tilde F)
\ee
 where $F_D=d V$.
 
\vskip .4cm
\noindent\hrulefill
\vskip .4cm

{\bf Exercise}: The action $S+\Delta S$ is quadratic in $F$. Integrate out $F$ to obtain the dual action:
\be
\tilde S={1\over 16\pi}Im\int \left(-{1\over \tau(a)}\right)(F_D^2+iF_D\tilde F_D)
\ee

\vskip .4cm
\noindent\hrulefill
\vskip .4cm

The above indicates that near the point where the monopole becomes massless 
the low energy theory contains the photon as well as the monopole.
By doing a duality transformation as above we can write the low energy theory in terms of the dual photon. With respect to it the monopole is electrically charged, and if the coupling is weak one can use normal perturbation theory.

We can choose a local coordinate $A(p)=C(u-u_0)$ around the point $u_0$ where the monopole becomes massless.
The mass of the monopole behaves as $M^2\sim |A(p)|^2$.
The theory around that point is IR free (since it is photons plus charges).
As we go go close to the singularity, $M\to 0$, perturbation theory (in the dual variables) becomes better and better.
The $\beta$-function coefficient due to a charged hypermultiplet is $b_H=
4{1\over 12}+4{1\over 6}=1$.
This implies that locally the prepotential is 
\be
\f=-{1\over 4\pi}A^2(p)\log{A^2(p)\over \tilde\Lambda^2}
\ee
and the dual coordinate
\be
A_D(p)\equiv {\partial \f\over \partial A(p)}=-{iA\over 2\pi}\left[\log{A^2(p)\over \tilde\Lambda^2}+1\right]
\label{ad}\ee
Now we can go around $u_0$: $u-u_0\to (u-u_0)e^{2\pi i}$.
Since $A(p)=C(u-u_0)$ we obtain that $A(p)\to A(p)$.
Also from (\ref{ad}) we obtain
$A_D(p)\to A_D(p)+2A(p)$.
Thus the monodromy matrix is
\be
\left(\matrix{A_D\cr A}\right)\to \hat M_{(0,1)}\left(\matrix{A_D\cr A}\right)=\left(\matrix{1&2\cr 0&1}\right)\left(\matrix{A_D\cr A}\right)
\ee
However we are interested in the monodromy matrix in the original variables.
We have performed a $\tau \to -1/\tau$ transformation in order to map the monopole to an electric charge.
We now have to invert this transformation.
We find
\be
 M_{(0,1)}=\left(\matrix{0&-1\cr 1& 0}\right)~\hat M_{(0,1)}~  \left(\matrix{0&1\cr -1& 0}\right)=\left(\matrix{1&0\cr -2& 1}\right)
\ee

\vskip .4cm
\noindent\hrulefill
\vskip .4cm

{\bf Exercise}: Consider a point where the $(n_e,n_m)$ dyon becomes massless.
By doing the appropriate duality transformation it can be treated as an electrically charged particle, whose local monodromy we have already computed. Invert the duality map to compute the monodromy matrix and show that 
\be
M_{(n_e,n_m)}=\left(\matrix{1-2n_en_m & 2 n_e^2\cr -2n_m^2 & 1+2n_e n_m\cr}\right)
\ee

\vskip .4cm
\noindent\hrulefill
\vskip .4cm

If the dyon (n,m) becomes massless at $u=\Lambda^2$ and ($n',m'$) at  $u=-\Lambda^2$ then we must have 
\be
M_{n,m}M_{n',m'}M_{\infty}=1
\ee
This can be solved to find the following solutions

\vskip 1cm
\centerline{
\begin{tabular}{|c|c|c|c|c|}\hline
(m,n)&(1,n)&(-1,n)&(-1,n)&(1,n)\\ \hline
(m',n')&(1,n-1)&(1,-n-1)&(-1,n+1)&(-1,-n+1)\\\hline
\end{tabular}}
\bigskip

The simplest solution is obtained for $m=m'=1$, $n=0,n'=-1$.
It can be shown that it is the only consistent solution.

So we are almost finished. We know all singular points of the 
holomorphic frame $(A(u),A_D(u))$ and the associated monodromy matrices.
It remains to use them to solve for $A(u), A_D(u)$.
The answer is that $A(u),A_D(u)$ are given by the two periods of an auxiliary torus.
The effective coupling constant $\tau$ is given by the modulus of the torus.
The periods of this torus vary as we change the modulus $u$.

The explicit solution can be written in terms of hypergeometric functions
\cite{sw}
\be
A(u)={\sqrt{2}\over \pi}\int_{-1}^1dx{\sqrt{x-u}\over \sqrt{x^2-1}}=
\sqrt{2(1+u)}~F\left(-{1\over 2},{1\over 2},1;{2\over 1+u}\right)
\label{sol1}\ee
\be
A_D(u)={\sqrt{2}\over \pi}\int^{u}_1dx{\sqrt{x-u}\over x^2-1}=
{i\over 2}(u-1)~F\left({1\over 2},{1\over 2},2;{1-u\over 2}\right)
\label{sol2}\ee
$F(a,b,c;x)$ is the standard hypergeometric function.
We have set $\Lambda=1$. It can be put back in on dimensional grounds.
Once we have (\ref{sol1},\ref{sol2}) we can compute the effective coupling $\tau$ as
\be
\tau (u)={A_D'\over A'}
\ee
where the prime stands for the $u$-derivative.
 
The positions of the three singularities coincide with the positions where the auxiliary torus degenerates (a cycle shrinks to zero).

In conclusion we have managed to calculate the exact low energy two-derivative effective action of an SU(2) N=2 gauge theory.
This theory has one parameter: the ultraviolet value of the coupling constant
or equivalently $\Lambda$.
For $|A|>>\Lambda$ the effective theory is weakly coupled and perturbation theory is reliable.
However, here we have controled the effective theory for $|A|\leq \Lambda$ 
where the effective coupling is strong.

The appearance of the torus in the Seiberg-Witten solution can be explained naturally by embedding the gauge theory into string theory \cite{lerche1}.

\section{Monopole condensation and confinement}
\setcounter{equation}{0}

Consider a U(1) gauge theory (QED) which is spontaneously broken by the 
non-zero vacuum expectation value of a (electrically charged) scalar field (Higgs). This is precisely what happens in normal superconductors.
The appropriate Higgs field is a bound state of electrons (Cooper pair)
with charge twice that of the electron.
Electric charge condenses in the vacuum (= the Higgs gets an expectation
value) and the photon becomes massive.

A well known phenomenon in such a phase is the Meissner effect.
Magnetic fields are expelled from the superconducting bulk.
There is only a thin surface penetration which goes to zero with the distance from the surface as $e^{-m~r}$. This is because the photon is massive in the superconductor and the parameter $m$ is no other than the photon mass.  
Thus, magnetic flux is screened inside a superconductor.

Consider now introducing a magnetic monopole inside the superconducting phase.
The magnetic flux emanating from the monopole will be strongly screened and will form a thin flux tube. If there is an anti-monopole around, the flux tube will stretch between the two.
At low energies such a flux tube is elastic and behaves like a string: the energy is proportional to the stretching.
Thus, there is a linear potential between a monopole-anti-monopole pair inside a superconductor. 
This means that magnetic mono\-poles are permanently confined in the superconducting (electric Higgs) phase.
As we try to pull them apart we must give more and more energy. Eventually 
when we have given energy greater than that required for a monopole-anti-monopole pair to materialize from the vacuum the string will break and we will end up with two bound states instead of separated magnetic charges.

The dual phenomenon was argued to be the explanation for the permanent confinement of quarks \cite{man}.
Here, we need a magnetically charged object (monopole) to get an expectation value in the vacuum (magnetic condensation).
The ensuing dual Meissner effect will confine the electric flux and the electric charges.
Although this mechanism remains to be seen if it is responsible for confinement in QCD, we will argue here following \cite{sw} that it does explain confinement in an N=1 gauge theory that we will obtain by perturbing the N=2 gauge theory we have considered so far.

We would like to softly break the original N=2 SU(2) gauge theory to N=1.
For this we split the N=2 vector multiplet into an N=1 vector multiplet $ (A^a_{\m}, \chi^a)$ and in a N=1 chiral multiplet $\Phi\equiv
(\psi^a, A^a)$.
We will add a superpotential ${\cal V}\sim m~Tr \Phi^2$ 
to make $\Phi$ massive.
At energies much smaller than $m$, $\Phi$ decouples and the theory is N=1 SU(2) super Yang-Mills which is an asymptotically free theory.
Thus, we would expect confinement, a mass gap and breaking of chiral symmetry
(which here is $\Z_4$ as discussed before).
 
 Consider the superpotential ${\cal V}=m~Tr \Phi^2/2=m~U$ where $U$ is the N=1 superfield whose scalar component is our coordinate $u$.
 If one goes through the same procedure of integrating out massive states
 one would get an extra potential in the low energy effective theory.
 It can be shown \cite{sw} that the induced superpotential is identical with the ultraviolet one.
 Consider now the effective theory near the point where the magnetic monopole becomes massless.
To smooth out the effective field theory we must include the monopole multiplet in our effective action.
The superpotential has an N=2  piece that gives the mass to the monopole $\sim |A_D|$ as well as the N=1 superpotential
\be
W=\sqrt{2}A_D~\tilde MM+m~U(A_D)
\ee
where $M,\tilde M$ denote the two N=1 components of the monopole 
hypermultiplet.
To find the ground state of the effective field theory we must minimize the potential: $dW=0$
\be
\sqrt{2}M\tilde M+m{du\over dA_D}=0\sp A_D~M=A_D~\tilde M=0
\label{vac}\ee

At a generic point $A_D\not= 0$ the solution to the second equation (\ref{vac}) is
$<M>=<\tilde M>=0$ .
Substituting in the first equation we obtain $du/dA_D=0$.
This can never be true since $u$ is a good global coordinate on the moduli space.

The only stable vacuum in the neighborhood exists for $A_D=0$.
From (\ref{vac}) we find that the monopoles have a non-trivial 
expectation value
\be
<M>=<\tilde M>=\sqrt{-{m\over \sqrt{2}}u'(0)}
\label{vac2}\ee
It can be checked from the exact solution that $u'(0)$ is negative.

What we have found is: a magnetically charged scalar has acquired a vacuum expectation value. It breaks the (magnetic) U(1) gauge group and generates confinement for the electric charges.
The fate of the massless fields is as follows: the U(1) vector multiplet
acquires a mass from the Higgs mechanism while the monopole hypermultiplet
is ``eaten up" by the vector multiplet.
The upshot  is that everything is now massive
and the  mass gap is proportional to the Higgs expectation value in 
(\ref{vac2}). This value is non-perturbative.

A similar analysis around the point where the dyon becomes massless gives similar results. There we have a realization of the oblique confinement of 't Hooft.
Thus, the theory we started with has two ground states, and this is explained by the chiral symmetry being broken from $\Z_4 \to \Z_2$.

\begin{figure}
\begin{center}
\vspace{0cm}  
\epsfig{file=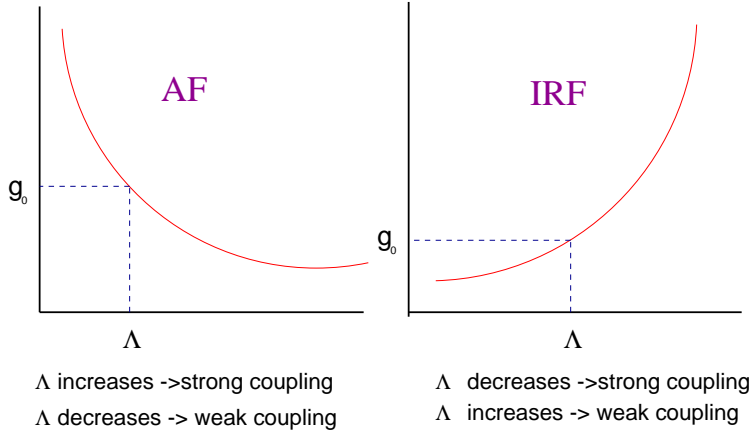,width=10cm}
\caption{Running coupling for asymptotically free and infrared free theories}
\label{fig3}
\end{center}
\end{figure}

\section{Epilogue of field theory duality}
\setcounter{equation}{0}

We have seen that in an N=4 supersymmetric field theory we expect an exact duality symmetry that interchanges weak with strong coupling.

In the context of N=2 gauge theories the solutions of Seiberg and Witten do generalize to arbitrary gauge groups\cite{gauge1} as well as the inclusion of ``matter" (hypermutiplets).
The exact effective description can always be found both in the Coulomb as well as in the Higgs phase.
There can be also mixed phases but they can be treated similarly.

The situation becomes more interesting in the context of N=1 gauge theories.
A general non-renormalizable N=1 field theory is specified by 
three functions of the 
chiral fields:

$\bullet$: The K\"ahler potential $K(z^i,\bar z^i)$ this is a real function and determines the kinetic terms of the chiral fields. Their geometry is that of a 
K\"ahler manifold with metric $G_{i\bar j}=\partial_i\partial_{\bar j}K$.

$\bullet$ The superpotential $W(z^i)$. It is a holomorphic function of the chiral fields and has R-charge equal to two.
The potential can be written in terms of the superpotential, and 
the K\"ahler metric (we ignore D-terms) as
\be
V\sim G^{i\bar j}\partial_i W\partial_{\bar j}\bar W
\ee

$\bullet$ The gauge coupling function $f(z)$.
It is also a holomorphic (and gauge invariant) function. Its imaginary part
determines the gauge coupling constant while its real part the $\theta$-angle.

In the N=1 case, unlike the N=2, we do not have full control over the 
two-derivative effective action.
We can determine however the holomorphic superpotential.
Assuming smoothness of the unknown K\"ahler potential, knowledge of the exact superpotential specifies uniquely the minima and thus the ground-states of the effective field theory.
Here again the strategy is to start from a renormalizable, 
asymptotically free gauge theory and find the superpotential in the low energy (strongly coupled ) effective field theory as well as the ground states. 

The N=1 SU($N_c$) gauge theory was studied \cite{sei} 
coupled to $N_F$ chiral multiplets 
in the fundamental and its complex conjugate.
We will briefly present some of the most interesting results. For more details
the interested reader should consults more extensive reviews on the subject
\cite{is} as well as the original papers \cite{sei}.

When $N_F>N_c+1$ the theory has a dual ``magnetic" description:
the dual gauge group is SU($N_F-N_c$) and the charged matter is composed of $N_F$
flavors of quarks as well as a set of  $N_F^2$ gauge singlet 
``mesons" $M_{ij}$.
These meson superfields are supposed to correspond to the mesons
of the original theory
\be
M_{ij}={1\over \mu}~q^i\bar q^j 
 \ee
 where $\mu$ is a dynamical scale.

Moreover there is an electric-magnetic type duality between the two theories (Seiberg duality) which can be expressed as a relationship between their $\Lambda$ parameters
as follows:
\be
\Lambda^{3N_c-N_F}~\tilde \Lambda^{3\tilde N_c-N_F}=(-1)^{N_c-N_F}\mu^{N_F}
\label{dual}\ee
where $\tilde N_c=N_F-N_c$.

The one-loop $\beta$-function coefficient of the original theory is 
$b=N_F-3N_c$ while that of the dual theory $\tilde b=3N_c-2N_F$.

In the range $N_c+1\leq N_F<{3\over 2}N_c$ the electric theory is asymptotically free while the magnetic theory is IR free.
Thus, the magnetic theory can be used to describe the low-energy dynamics in a weak coupling regime.
The relation (\ref{dual}) can be seen to indicate that when the electric coupling is strong the magnetic coupling is weak and vice versa (see Fig. 
\ref{fig3}).   
 In the region ${3\over 2}N_c\leq N_F\leq 3N_c$ both theories are AF and they 
flow to a non-trivial fixed point in the IR.

An interesting and important question is: what can be done when there is no supersymmetry or when supersymmetry is broken?
Duality ideas seem that they can handle the softly broken case \cite{ak}.
However, calculations in the  broken theory can be trusted once the supersymmetry breaking scale is much smaller that the dynamical scale(s) of the theory.

\begin{figure}
\begin{center}
\vspace{0cm}  
\epsfig{file=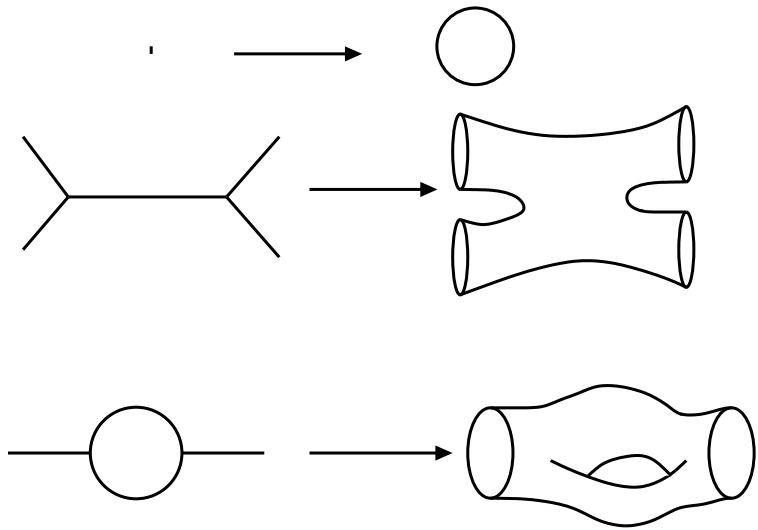,width=10cm}
\caption{String theory versus field theory diagrams}
\label{fig4}
\end{center}
\end{figure}

\section{Introduction to String Theory}
\setcounter{equation}{0}

String theory was born in 1968 \cite{ven} 
as a candidate theory to describe the dual properties of hadrons.
It has been superseded by QCD, and reemerged in 1976 \cite{ss} as a candidate theory of gravity and all other fundamental interactions.
In 1984 it acquired a big impetus \cite{gs} due to the tightness of constraints
\cite{aw} on possible consistent theories.

String theory postulates that the fundamental entities are strings rather than point-like objects. However from a large distance a string can be viewed as a point-like object. Thus, at distances well above the string length $l_s$
string theory is well approximated by field theory.
String perturbation theory resembles field theory perturbation theory, (diagrams
fatten, see Fig. \ref{fig4}) but has also different properties in the UV.

\begin{figure}
\begin{center}
\vspace{0cm}  
\epsfig{file=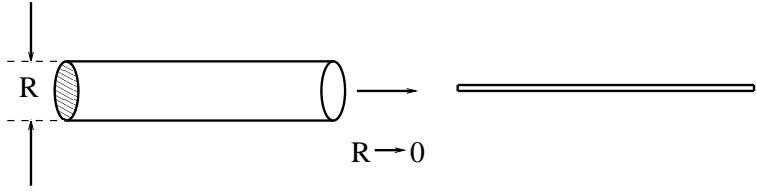,width=10cm}
\caption{A circular extra dimension can be invisible when $R$ is small.}
\label{fig5}
\end{center}
\end{figure}

$\bullet$ Closed string theory predicts gravity. If one quantizes free strings in a flat background, a spin-two massless state can be found in the spectrum.
It has gravitational type interactions and can be identified with the graviton.

$\bullet$ It is a theory that is UV-finite. In some sense string theory can be though of as a collection of an infinite number of quantum fields with a
``smart" UV cutoff of the order of the string scale $M_s$.

$\bullet$ String theory provides a consistent and finite theory 
of perturbative quantum gravity.

$\bullet$ Existence of space-time fermions in the theory implies supersymmetry.

$\bullet$ String theory unifies gravity with gauge and Yukawa interactions
naturally.

$\bullet$ The theory has  no free parameters (apart from a scale $l_s=M_s^{-1}$) but many ground-states (vacua). The string coupling constant is related to the expectation value of a scalar field, the dilaton $g_s=e^{<\phi>}$.
The continuous parameters of various ground-states are always related
to expectation values of scalar fields.
The string tension is $l_s^{-2}$.

$\bullet$ String theory was defined in perturbation theory until '95.
Since then duality ideas allowed us to explore string theory 
beyond perturbation theory, indicate that the theory is unique, suggest the existence of a most symmetric eleven-dimensional theory and provided new tools and ingredients for its study. We do not as of now have a complete non-perturbative formulation of the theory.

$\bullet$ Superstrings live in ten or less large (non-compact) dimensions.
A topical question is: How come we see four large dimensions today?
Kaluza and Klein long time ago, suggested how the two could be compatible.
The idea is that some dimensions can be small and compact and can thus avoid detection (see Fig. \ref{fig5}).
We will present here a five-dimensional example for the sake of simplicity.
We consider a massless scalar in five dimensions, with mass-shell condition 
$p^2=0$.
Consider now the fifth coordinate to be a circle with radius R.
Then the components of the momentum along the fifth directions is quantized. This is obtained from the periodicity of the wavefunction $e^{ip_5\;x^5}$
under shifts $x^5\to x^5+2pi R$.
One obtains $p_5={m\over R}$ where $m\in \Z$.
Thus, the five-dimensional mass-shell condition can be written as 
\be
p_0^2-\vec p^2={m^2\over R^2}
\label{KK}\ee
Equation (\ref{KK}) indicates that from the four-dimensional point of view, 
this five-dimensional massless scalar corresponds to an infinite tower of particles (called Kaluza-Klein states) with masses $M={|m|\over R}$.
When our available energy $E<<1/R$ no experiment can produce a KK particle.
Moreover, their loop effects are suppressed.
Thus, at $E<<1/R$ the extra dimension is unobservable.
For $E\geq 1/R$ the effects of the KK particles and thus the extra dimension become visible.
For example, if such a particle feels the standard model forces then this implies an upper bound on the radius which is of the order $R \sim 10^{-20} m\sim (10 TeV)^{-1}$ \cite{anto1}.
On the other hand it is quite surprising that if only gravity lives in five dimensions, then the radius can be as large as $R\sim 10^{-4} m$ \cite{dim1}
without contradicting current experimental data \cite{dim2}.

The ten-dimensional part of the action governing low energy gravity (below the string scale) is
\be
S_{10}\sim {1\over g_s^2l_s^8}\int d^{10}x~\sqrt{-G}~R+\cdots
\label{10d}\ee
If six dimensions are compactified on a manifold with volume $V_6l_s^6$ then the four-dimensional Einstein action obtained from (\ref{10d}) will be
\be
S_4\sim {V_6\over g_s^2 l_s^2}  \int d^{4}x~\sqrt{-g}~R+\cdots
\ee
from where we can read the four-dimensional Planck mass
\be
M_P^2 \simeq (10^{19} GeV)^2= {V_6\over g_s^2 l_s^2}\spa\Rightarrow \spa{M_s\over M_P}={g_s\over \sqrt{V_6}}
\label{mp} \ee
The following regimes are important:

$\bullet$ For energies below the string tension, $E<M_s$, strings cannot have their vibrational modes excited. Their dynamics is associated with their center of mass motion and can be thus described by standard field theory.
On the contrary, for $E>M_s$ the stringy modes can be excited and the physics
departs sensibly from the field theory behavior.

$\bullet$ For energies $E<<M_P$ gravity is very weak and can be neglected at the microscopic level. Its quantum effects are unimportant.
On the other hand, for $E\geq M_P$ gravity becomes strong, th quantum gravitational effects cannot be treated perturbatively and it is not 
known how to handle the theory in this case.

There are three possibilities concerning the hierarchy of scales:

\begin{itemize}

\item $M_{P}\sim M_s$. This is the conventional scenario, where gauge fields come from the perturbative closed string sector.
Both stringy as well as quantum gravitational phenomena are far removed from near future experiments and experimental signals of string theory are obscured by the huge disparity in scales.
For this to happen,  $g_s\sim {\cal O}(\sqrt{V_6})$. 
There are two possibilities in perturbation theory: $g_s \leq {\cal O}(1)$ and a compact manifold of Planck size, or an hierarchically small coupling constant
and a sub-Planckian compact manifold.
In the second case, this can be mapped via T-duality to a string theory with a large volume compact manifold. 

\item $M_s<< M_P$. In this case stringy phenomena can be visible at low energy, hopefully at near future accelerator experiments, while quantum gravity
 remains out of current reach. For this to work out, there are two possibilities.
 First, a hierarchically small coupling constant and a Planckian size manifold.
 Second, $g_s \leq {\cal O}(1)$ and a large compact manifold.
 In this case the threshold of KK excitations is of order $M_{KK}\sim {M_s\over V_6^{1/6}}<<M_s$. Thus the first signal will be production of KK states before stringy effects are visible.
 
 \item $M_P<<M_s$. This necessitates (in perturbation theory) a sub-Planckian compact space which will be mapped via T-duality to a different string ground state.

\end{itemize}

\section{T-duality}
\setcounter{equation}{0}

Classical strings behave very different from point particles at distances of order the string length, $l_s$.
A characteristic feature is that closed strings can stretch and wrap 
around a non-contractible cycle of a compact manifold.
Consider again the five-dimensional example with one direction being a circle of radius R.
The energy cost for a string wrapping n times around the circle is given by
\be
E_{\rm wrapping}=({\rm total~~ length})\times ({\rm string~~ tension})=2\pi n{R\over l_s^2}
\ee
Now, the mass-shell condition (\ref{KK}) is modified to
\be
p_0^2-\vec p^2={m^2\over R^2}+4\pi^2R^2{n^2\over l_s^4}
\label{kk1}\ee

A symmetry (a special case of $T-duality$) is obvious in the mass formula (\ref{kk1}):
\be
R\to \tilde R={l_s^2\over 2\pi R}\sp m\leftrightarrow n
\label{td}\ee 
The physical content of this stringy symmetry is that we cannot distinguish a circle with size smaller than the string length.
The effective radius we measure is always 
\be
R_{\rm eff}\geq {l_s\over \sqrt{2\pi}}
\ee
When R is large the low lying excitations are the KK states. When R is small, the low lying excitations are winding modes, that can be interpreted as KK modes 
with a dual radius $\tilde R$.
T-duality is a symmetry of string theory valid order by order in perturbation theory. 

The fact that the string cannot distinguish length scales that are smaller that its size is no surprise. What is a surprise is that a circle with length much smaller than the string length is equivalent to a macroscopic one.

Classical strings at distances larger than the string scale, feel the standard Riemannian geometry. At smaller scales, the Riemannian concept breaks down.
The generalization is provided by Conformal Field Theory which could be viewed
as an infinite-dimensional generalization of Riemannian geometry \cite{k1,k2}.
This can have deep implications on the geometric interpretation of strong curvature as well as early cosmological phenomena \cite{k2}.

\section{A collection of superstring theories}
\setcounter{equation}{0}

Until recently we were blessed with an embarassement of riches: we knew
five distinct, stable, consistent, supersymmetric string theories in ten dimensions.

\underline{Closed Strings}

$\bullet$ {\bf Type-II strings}. These are the most normal of all strings.
They are closed strings, with isomorphic left-moving and right-moving modes. There are also fermionic oscillations responsible for the appearance
of space-time fermions.
They are Lorentz invariant in ten-dimensional flat space.
There is a subtle difference of ``gluing" together the fermionic left and right movers. This results in two distinct string theories:

\begin{itemize}

\item {\bf type IIA}: This is a non-chiral ten-dimensional theory with N=2 space-time supersymmetry.
The low energy effective field theory is type IIA supergravity.
Its bosonic spectrum  contains the graviton, a
 two-index antisymmetric tensor and a scalar (the dilaton) as well
as a set of forms (Ramond-Ramond states): a vector and a three-form.

\item {\bf type IIB}: This is a chiral, anomaly-free ten-dimensional theory 
with N=2 supersymmetry. The low energy effective field theory is type-IIB
supergravity.
The bosonic spectrum contains the graviton, two-form and dilaton (like the type IIA) but the Ramond-Ramond (RR) forms are different: here we have a zero-form (scalar), another two-form and a self-dual four-form.
One can make a complex number $\tau=a+ie^{-\phi}$, by putting together the 
RR scalar (axion, a) and the dilaton (string coupling constant, $g_s=e^{\phi}$)
Then, the effective type-IIB supergravity is invariant under a continuous 
SL(2,R) symmetry which acts projectively on $\tau$:
\be
\left(\matrix{a&b\cr c&d}\right)\in {\rm SL(2,R)}\sp \tau\to {a\tau+b\over c\tau+d}
\ee
The two two-forms transform as a doublet, while the Einstein metric and the four-form are invariant.
This is reminiscent of a similar situation in $N=4$ super Yang-Mills theory.
It is expected that the presence of objects charged under the two-forms will 
break the continuous symmetry to a discrete subgroup, namely SL(2,$\Z$).

\end{itemize} 
Both type II strings cannot fit the fields of the Standard Model in $perturbation$ $theory$. This is partly due to the fact that gauge fields descending from the RR sector have no charged states in perturbation theory
and cannot thus serve as Standard Model gauge fields.

\begin{figure}
\begin{center}
\vspace{0cm}  
\epsfig{file=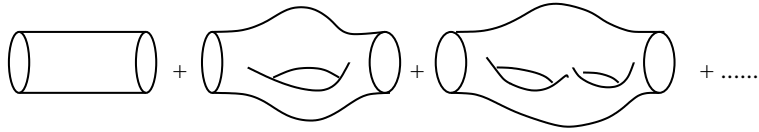,width=10cm}
\caption{The first few diagrams for the propagator of a closed string theory}
\label{fig6}
\end{center}
\end{figure}

\begin{figure}
\begin{center}
\vspace{0cm}  
\epsfig{file=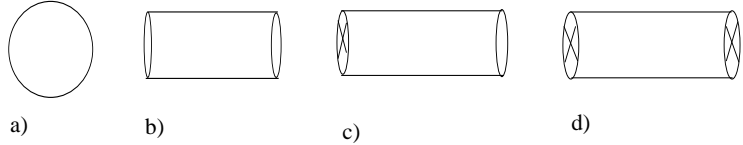,width=10cm}
\caption{The first few diagrams (with boundaries and unorientable surfaces)  
for the vacuum energy of an open string theory: (a) Disk (b) Annulus (c) Moebius strip  (d) Klein bottle}
\label{fig7}
\end{center}
\end{figure}

$\bullet$ {\bf Heterotic strings}. This is a peculiar type of string \cite{gro}. The idea is that since left and right-movers are independent one can glue  superstring modes on the right (living in ten dimensions) and bosonic string modes on the left (living in twenty six dimensions).
The extra sixteen left-moving coordinates are required by consistency to be compactified on the two possible even self-dual sixteen-dimensional lattices:
the root lattice of E$_8\times$E$_8$ or that of Spin(32)/$Z_2$\footnote{This is the root lattice of SO(32) augmented by one of the two spinor weights.}.
The low energy effective field theory is N=1 D=10 supergravity coupled to D=10 super Yang-Mills with gauge group E$_8\times$E$_8$ or SO(32).
The bosonic spectrum is composed of the metric two-form and dilaton, as well as 
the gauge bosons in the adjoint of the gauge group.

For all the closed string theories the structure of perturbation theory is
elegant: each order of perturbation theory corresponds to a computation 
using the appropriate Conformal Field Theory on the associated Riemann surface.
The perturbative expansion is organized by the number of loops (genus or number of handles of the associated Riemann surface), and there is a single diagram per order. This includes (in the low energy limit) the contributions of $N!$
distinct diagrams of field theory (see Fig. \ref{fig6}).  

\underline{Open and Closed Strings}: {\bf Type-I} string theory.
The theory contains both closed and open unoriented strings.
From the closed string sector we obtain N=1 supergravity in ten dimensions
while from the open string sector we obtain SO(32) super Yang-Mills.
The structure of the perturbation theory is more involved now since it involves both open and closed surfaces, as well as both orientable and non-orientable
surfaces. The first few extra terms in the genus expansion are shown in Fig. \ref{fig7}.

\begin{figure}
\begin{center}
\vspace{0cm}  
\epsfig{file=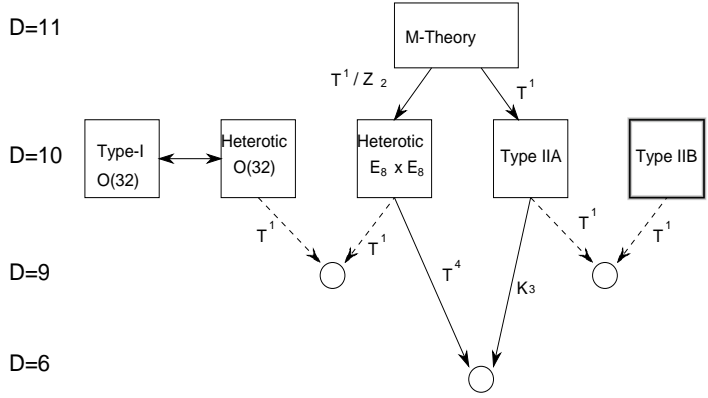,width=10cm}
\caption{Perturbative and non-perturbative connections between string theories}
\label{fig8}
\end{center}
\end{figure}

\section{Duality connections}
\setcounter{equation}{0} 

We have seen that we have five distinct supersymmetric theories in ten dimensions. Are they truly distinct or they form part of an underlying theory?

In string perturbation theory there are two connections that are shown in Fig. \ref{fig8} with broken arrows.

\begin{figure}
\begin{center}
\vspace{0cm}  
\epsfig{file=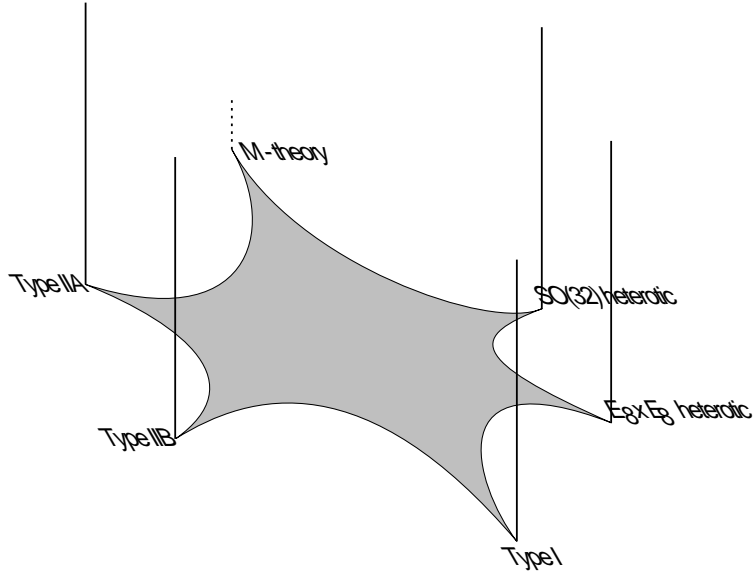,width=10cm}
\caption{Perturbative and non-perturbative connections between string theories}
\label{fig9}
\end{center}
\end{figure}

Upon compactification to nine (or less) dimensions on a circle of radius $R$
the heterotic E$_8\times$E$_8$ and O(32) theories are continuously connected.
In nine dimensions, we can turn-on Higgs expectation values\footnote{These are scalars that come from the tenth components of the gauge fields in ten dimensions. These expectation values are called Wilson lines.}
and break the gauge group.
We have two limits in which we can go back to ten dimensions: The first is to take $R\to \infty$. If we started with the O(32) string we will end up with the O(32) string in ten dimensions.
The other is $R\to 0$. You remember that using T-duality $R\to 0$ is still equivalent to a very large circle. If we adjust appropriately the Wilson lines in this limit we end up with the  E$_8\times$E$_8$ string.
This indicated that the two ten-dimensional theories are not disconnected 
but corners in the same moduli space of vacua of a single (the heterotic) theory (see figure \ref{fig9}).

A similar situation exists for the type IIA and type IIB theories.
Although they look very different (for example one is chiral the other is not)
once they are compactified to nine dimensions they are related by T-duality.
Thus at $R=\infty$ one recovers the ten-dimensional type IIA theory while at 
$R=0$ we recover the ten-dimensional type IIB theory (figure \ref{fig9}). 

If we go beyond perturbation theory we will find more connections \cite{ht,wit1}. The key is to ask what is the strong coupling limit of the various ten-dimensional 
string theories. The tools to investigate this question we have already discussed in the field theory context: they are supersymmetry and BPS states.

\begin{figure}
\begin{center}
\vspace{0cm}  
\epsfig{file=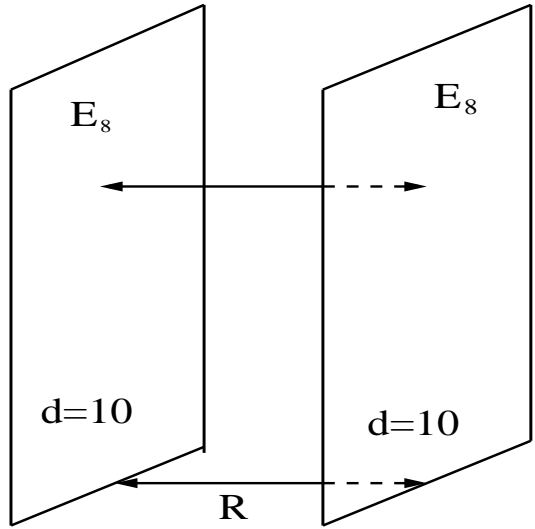,width=7cm,height=7cm}
\caption{The non-perturbative E$_8\times$E$_8$ heterotic string as a compactification of M-theory on a interval}
\label{fig10}
\end{center}
\end{figure}

$\bullet$ The type-IIA theory contains point-like solitons (known today as $D0-branes$) that are electrically charged under the RR gauge field (remember no perturbative state has electric or magnetic charge under RR forms).
Their mass is given by
\be
M_{D0}={n\over g_s}\sp n\in \Z
\label{d0}\ee
where $n$ is the electric charge.
Since these are 1/2-BPS states we can trust their mass formula also at strong coupling. Thus, we learn that at strong coupling they become arbitrarily light. This tower of states reminds us of the tower of KK states for large radius.
This is not accidental: it was long known that the action of ten-dimensional
type-IIA supergravity could be obtained by dimensional reduction of eleven-dimensional supergravity on a circle of radius $R$.
The KK states of the graviton  have a spectrum like the one in (\ref{d0})
and they are charged under the off-diagonal components of the 
eleven-dimensional metric that becomes the RR gauge field.
The precise relation is 
\be
g_s=R^{2/3}
\label{m}\ee
Thus, we expect that the strong coupling limit of type IIA theory is an eleven-dimensional theory (named M-theory)  whose low-energy limit is eleven-dimensional supergravity \cite{wit1}.
Compactifying M-theory on a circle we obtain type-IIA string theory.

$\bullet$ On the other hand compactifying M-theory on the orbifold $S^1/\Z_2$ we obtain the E$_8\times$E$_8$ heterotic string theory.
The string coupling and the radius of the orbifold are still related as in (\ref{m}).
The orbifold is defined by moding the circle out by the inversion of the coordinate $\sigma \to -\sigma$.
This projects out the low energy spectrum of M-theory to N=1 ten-dimensional supergravity. 
We also have two fixed points of the action of the orbifold transformations: $\s=0,\pi$. These are fixed ten-dimensional planes, and as it happens in 
perturbative string theory, there are extra excitations localized on the orbifold planes. Anomaly cancellation indicated that each plane should carry a ten-dimensional E$_8$ Yang Mills supermultiplet (figure \ref{fig10}).
Thus, in the perturbative heterotic string (small R) the two planes are on top of each other whereas they move apart non-perturbatively.

$\bullet$ The strong coupling limit of type-IIB theory is isomorphic to its weak coupling limit. This is due to the fact that an SL(2,$\Z$) subgroup of 
the continuous SL(2,R) symmetry is unbroken and that includes the transformation that inverts the coupling constant.

$\bullet$ Finally, the two O(32) theories, namely the heterotic and the type-I are dual to each other. This means that the strong coupling limit of the heterotic theory is the weakly coupled type-I theory and vice versa.

All these connections are summarized in figure \ref{fig10} and the overall picture is portrayed in figure \ref{fig9}.
We learn that the five string theories are corners in a moduli space of a 
more fundamental theory.

\section{Forms, branes and duality}
\setcounter{equation}{0}

We have seen that the various string theories have massless
antisymmetric tensors in their spectrum. We will describe here 
 the natural
charged objects of such forms and how electric-magnetic 
duality extends to them.

We will use the language of differential forms and we will represent a rank-p
antisymmetric tensor $A_{\m_1\m_2\ldots\m_p}$ by the associated p-form
\be
A_p\equiv
A_{\m_1\m_2\ldots\m_p}dx^{\m_1}\wedge\ldots\wedge dx^{\m_p}
\,.\label{562}\ee
Such p-forms transform under generalized gauge transformations:
\be
A_p\to A_p+d~\Lambda_{p-1},
\,,\label{563}\ee
where $d$ is the exterior derivative ($d^2=0$) and $\Lambda_{p-1}$ is a
$(p-1)$-form that serves as the parameter of gauge transformations.
The familiar case of (abelian) gauge fields corresponds to p=1.
The gauge-invariant field strength is
\be
F_{p+1}=d~A_{p}
\,.\label{564}\ee
satisfying the free Maxwell equations
\be
d^{*}F_{p+1}=0
\label{5644}\ee

The natural objects, charged under a (p+1)-form $A_{p+1}$, are
$p$-branes.
A $p$-brane is an extended object with $p$ spatial dimensions.
The world-volume of $p$-brane is (p+1)-dimensional.
Point particles correspond to p=0, strings to p=1.
The natural coupling of $A_{p+1}$ and a p-brane is given by
\be
\exp\left[iQ_p\int_{\rm world-volume} A_{p+1}\right]=
\exp\left[iQ_p\int A_{\m_0\ldots\m_p}dx^{\m_0}\wedge\ldots\wedge
dx^{\m_p}\right]
\,,\label{565}\ee
which generalizes the Wilson line coupling in the case of
electromagnetism.
This is the $\s$-model coupling of the usual
string
to the two-index antisymmetric tensor.
The charge $Q_p$ is the usual electric charge for p=0 and the string
tension for p=1. $Q_p$ has mass dimension $p+1$.
For the p-branes we will be considering, the (electric) charges will be
related to their tensions (mass per unit volume).

In analogy with electromagnetism, we can also introduce magnetic
charges.
First, we must define the analog of the magnetic field: the magnetic
(dual) form.
This is done by first dualizing the field strength and then rewriting
it as the exterior derivative of another form\footnote{This is
guaranteed by (\ref{5644}).} :
\be
d\tilde A_{D-p-3}=\tilde F_{D-p-2}=^*F_{p+2}=^*dA_{p+1}
\,,\label{566}\ee
where D is the the dimension of space-time.
Thus, the dual (magnetic) form couples to $(D-p-4)$-branes that play
the role of magnetic monopoles with ``magnetic charges" $\tilde
Q_{D-p-4}$.

There is a generalization of the Dirac quantization condition to
general
p-form charges discovered by Nepomechie and Teitelboim \cite{nep}.
The argument parallels that of Dirac. Consider an electric p-brane
with charge $Q_p$ and a magnetic $(D-p-4)$-brane with charge $\tilde
Q_{D-p-4}$.
Normalize the forms so that the kinetic term is ${1\over 2}\int
^*F_{p+2}F_{p+2}$.
Integrating the field strength $F_{p+2}$ on a (D-p-2)-sphere
surrounding
the p-brane we obtain the total flux $\Phi=Q_p$.
We can also write
\be
\Phi=\int_{S^{D-p-2}}~^*F_{p+2}=\int_{S^{D-p-3}}~\tilde A_{D-p-3}
\,,\label{567}\ee
where we have used (\ref{566}) and we have integrated around the
``Dirac string".
When the magnetic brane circles the Dirac string it picks up a phase
$e^{i\Phi\tilde Q_{D-p-4}}$, as can be seen from (\ref{565}).
Unobservability of the string implies the Dirac-Nepomechie-Teitelboim
quantization condition
\be
\Phi \tilde Q_{D-p-4}=Q_{p}\tilde Q_{D-p-4}=2\pi N\;\;\;,\;\;\;n\in \Z
\,.\label{568}\ee

The type IIA string theory contains a one- and a three-form in the RR sector.
They couple electrically to a particle (D0-brane) a membrane (D2-brane)
and magnetically to a D6-brane and and D4-brane. Moreover there is a non-propagating nine-form that couples to a D8-brane.
There is always the fundamental string that couples electrically to the two-index antisymmetric tensor. Its magnetic dual is the NS5-brane.

In the type IIB theory we have a zero- two- and self-dual four-form.
The electric and magnetic branes are the D1, D3, D5 and D7 branes. The is also a D-instanton (denoted also by D(-1)).

These branes can be described as solitonic extended objects in the low energy supergravity theory. All are 1/2-BPS states and thus preserve half of the original supersymmetry.

\begin{figure}
\begin{center}
\vspace{0cm}  
\epsfig{file=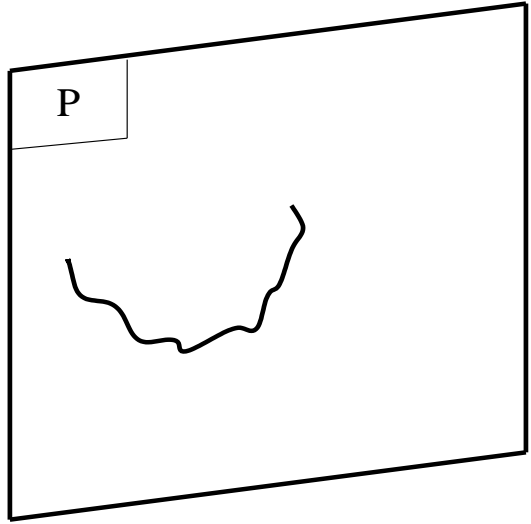,width=7cm,height=7cm}
\caption{A D-brane with an attached string}
\label{fig11}
\end{center}
\end{figure}

\section{D-branes}
\setcounter{equation}{0}
 
There is an exact stringy description of the solitonic branes we have mentioned in the previous section (except the NS5-brane). They can be defined as defects (walls) in space-time where closed strings can end. A closed string when moving always stays closed.
When it interacts with a brane it can open up and its end-points are forced to move on the brane (figure \ref{fig11}).
The fluctuations of such open strings are essentially the fluctuations of the brane itself.
They can be shown to carry the appropriate RR charge. Their name derives from the Dirichlet boundary conditions obeyed by the open strings attached to the brane.

The quantization of the open strings on a Dp-brane gives a massless spectrum that is that of
maximal Yang-Mills supermultiplet in p+1 dimensions.
It contains a single vector, 9-p scalars and the associated fermions.
Note that is is the dimensional reduction of an N=1 Yang-Mills supermultiplet in ten dimensions.

The p-brane has some obvious collective coordinates, namely its position in the 
transverse (9-p)-dimensional space. The expectation value of the 9-p scalars 
are precisely these collective coordinates. They have no potential since we can put a brane anywhere in the transverse space.
There is an effective action on the D-brane that describes its dynamics.
It can be calculated from the string description.
Since the D-brane is a 1/2-BPS state, its world-volume action will be supersymmetric (N=1 in ten dimensions (D9), or N=4 in four dimensions (D3)
etc.)
Moreover at low energies the action must reduce to the super Yang-Mills action.
The effective action is
\be
S_{eff}={1\over g_s}\int d^{p+1} x~\sqrt{det(g_{ab}-F_{ab})}+\cdots
\simeq 
\ee
$$
\simeq\int d^{p+1} x~e^{-\phi}\left[1+F_{ab}F^{ab}+\partial_{a}X^I\partial^{a}X^I+\cdots\right]
$$
where the induced metric on the brane is
\be
g_{ab}=\delta_{ab}+\partial_a X^I\partial_b X^I
\label{ind}\ee
This action describes in general the dynamics of the brane modes as well as 
their coupling  to the bulk string fields.
It is non-linear and comes under  the name of Dirac-Born-Infeld (DBI) action. 
Note that the energy per unit volume is proportional to $1/g_s$.
For normal solitons the dependence is $1/g^2$.

\begin{figure}
\begin{center}
\vspace{0cm}  
\epsfig{file=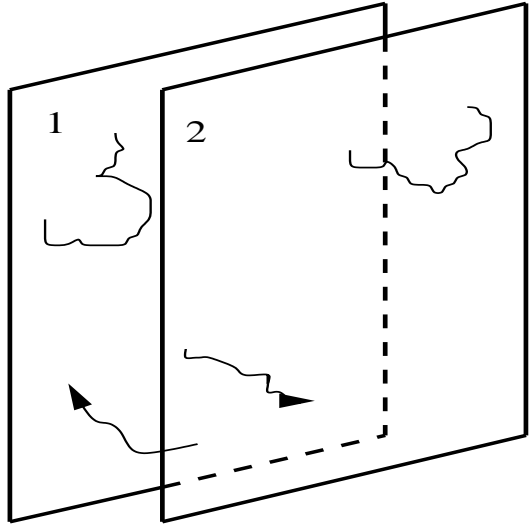,width=7cm,height=7cm}
\caption{Two parallel  D-branes and the various open string fluctuations}
\label{fig12}
\end{center}
\end{figure}

An interesting phenomenon happens when we have many coincident D-branes.
As can be seen in figure \ref{fig12}, if we label the branes by 1 and 2
then there are four possible strings:1-1, 2-2, 1-2, 2-1.
Each will give rise to a massless Yang-Mills multiplet (if the branes coincide in transverse space).
It turns out that the gauge symmetry now is non-abelian, namely U(2). 
This can be inferred from the fact that a string end-point on the 
brane acts like an electric charge for the
gauge field coming from the string with both end-points on the brane.
Note that the scalars $X^I$ that we had interpreted as the coordinates of the D-brane have now become $2\times 2$ matrices.
This is an interesting realization of ideas concerning the quantization of space-time (the coordinates becoming non-commuting operators).

What happens when by keeping the branes parallel we separate them a distance $l$ in the transverse space (figure \ref{fig12})?
The two strings (1-2, 2-1) are now stretched by a distance l and give a shift in the energy $\s~l$ where $\s$ is the string tension.
Thus, the two gauge bosons associated to them are no longer massless: they have
a mass $\s~l$.
In the effective theory on the branes, this is the ordinary Higgs effect.
The U(2) Yang Mills has a potential 
$$
V=Tr([X_I,X^{\dagger}_J][X_I,X^{\dagger}_J])
$$
The minimum is when $X^I$ are diagonal matrices (the Cartan of U(2))
\be
X^I_{min}=\left(\matrix{x^I_1&0\cr 0&x_2^I}\right)
\ee  
The vacuum expectation values $x^I_1,x_2^I$ have the interpretation of the coordinates of the transverse position of the two branes.
The two off-diagonal gauge fields acquire a mass proportional to $|x_1-x_2|$
in accordance with our expectations.

The generalization is straightforward for N parallel branes. The gauge group is U(N). The overall U(1) corresponds to the center of mass position while the SU(N) describes the internal dynamics.
In the generic vacuum the branes are all separate and the gauge symmetry is broken to $U(1)^N$.

The state of affairs has some important messages

$\bullet$ The space-time positions of branes correspond to the vacua of the world-volume Yang-Mills theory.

$\bullet$ The fluctuations of the D-branes are the fluctuations of the Yang-Mills theory.

$\bullet$ The interaction of the brane with the bulk supergravity fields is provided by the word-volume couplings.
For example the interaction with the space-time metric $G_{\m\n}$ is obtained from the following modification of the induced metric in 
(\ref{ind})
\be
g_{ab}=G_{\m\n}\partial_a X^{\m}\partial_b X^{\n}
\ee
etc.

There are numerous applications of the previous observations:

$\bullet$ Geometric/brane Engineering. The strategy here is the following.
We put together branes so that we construct our favorite gauge theory including some matter content.
We compactify some directions and take $g_s\to 0$ to decouple gravity while keeping the gauge coupling fixed.
We can then study properties of the associated Yang-Mills theories from the space-time picture of the branes.
The results include  a derivation of the M-O duality for N=4 super Yang-Mills theory, derivation of Seiberg-Witten type solutions for N=2 theories as well as the Seiberg Duality for N=1 gauge theories.

$\bullet$ Black-hole state counting.

$\bullet$ Gauge theory/gravity correspondence.

\section{Black-holes and D-branes}
\setcounter{equation}{0}

In the early seventies, culminating with the works of Bekenstein and Hawking
it was realized that black-holes obey laws similar to ordinary thermodynamics.

$\bullet$ They have entropy given by one quarter the area of the horizon in gravitational units
\be
S_{BH}={1\over 4}{A\over G_N}
\label{beke}\ee

$\bullet$ They radiate thermal radiation with temperature
\be
T_H={\kappa\over 2\pi}
\ee 
where $\kappa$ is the specific gravity on the horizon.

$\bullet$ They satisfy all thermodynamic laws. The first
\be
dM=TdS+{\rm work}
\ee
where work terms  can be related to angular momentum, charge etc.
The second $dS\geq 0$ is also satisfied (by classical gravity) 
as well as the third.

The above observations create a clash with quantum mechanics known as the 
``black hole information paradox" that can be summarized as follows:
Form a black hole from matter in an initially pure state.
Let it evaporate completely via thermal Hawking radiation.
Then the whole system has transformed into a mixed state and this is 
is not permitted by quantum mechanics.
There have been many attempts to resolve this paradox till today, but it is 
fair to say that the paradox still stands.

One can cook up a similar paradox with a star.
The star is formed by matter in a pure state that is eventually squeezed by gravity, heating up, and radiating thermal radiation.
Here however there is no paradox. We do know that the initial correlations are encoded in the outgoing radiation which is not exactly (only approximately)
thermal.
In order to argue this, we have as a tool the microscopic 
statistical mechanics of all particles that form a star.
Without knowing the microscopic degrees of freedom one cannot resolve this paradox.

Thus, the important question is: what are the microscopic degrees of freedom
responsible for the Bekenstein entropy (\ref{beke}) of a black hole?

Until a couple of years ago this question went unanswered.
Here we will show that string theory gives the microscopic degrees of freedom
responsible for the entropy.

Consider a particle of mass M. If the Schwarschild radius of the particle is much bigger than the fundamental gravitational length $l_P$ then the particle can be viewed as a black-hole.
This will happen if the particle has a mass much larger than the Planck mass.

String theory has many such states with masses $M>>M_P$. Moreover for large $M$
their density grows exponentially as $e^{c~M}$. This implies that their entropy is linear with the mass.
However, the Bekenstein entropy for black holes grows quadratically with the mass (in four dimensions).

There is already a problem with the comparison though.
The masses of string states obtain generically large quantum corrections.
The mass entering the Bekenstein formula is the physical mass (after quantum corrections have been taken into account) whereas the mass in the density is the bare mass.  
The way out is to look for states that are protected from quantum corrections. These are precisely the BPS states in supersymmetric theories.
Thus, we would like to put together many of those and create a smooth black hole.

  Consider a charged black hole, with charge Q.
  Then the BPS bound is $M\geq |Q|$. If the black hole is extremal (BPS) then $M=|Q|$ and it has zero Hawking temperature. It is stable as expected from supersymmetry.
It must however have a horizon of finite area (or equivalently, finite macroscopic entropy)    
  
The simplest example of that sort can be constructed in five dimensions in the type IIB theory.
In order to have an extremal black-hole with non-zero horizon area, it must carry three distinct charges, $Q_1$, $Q_5$ and $Q_0$ (in four dimensions
we need four).

The first is to find the supergravity solution, compute the area and then the Bekenstein entropy. This can be done and the result is
\be
S_{\rm Bekenstein}=2\pi\sqrt{Q_0Q_1Q_5}
\label{bek}\ee  

The second step is to construct the black-hole out of a collection of elementary states, in all possible ways, matching its charges.
We need to use D-branes (wrapped on cycles so as to give point-like objects in D=5).
So we consider type-IIB theory compactified on $C_4\times S^1$ (where $C_4$ can be $T^4$ or K3),  to five dimensions.
Consider a bound state formed out of $Q_5$ D5-branes compactified around
$C_4\times S^1$ and $Q_1$ D1 strings wrapped around $S_1$.
If we consider the volume of $C_4$ to be much smaller than that of $S^1$, the world-volume theory on the branes can be reduced to 1+1 dimensions (time+$S^1$).
We can still add some fluctuations without breaking supersymmetry.
 We can consider left-moving waves with ``energy", $Q_0$ in the (1+1)-dimensional theory. They do not break all of supersymmetry so we are still considering
an extremal configuration.
It can be shown that the numbers $Q_0,Q_1,Q_5$ correspond to gauge charges of the bound-state.

Now we would like to take the string coupling to be small $g_s<<1$ so that gravity is weakly coupled. We would like also to have a bound-state that is macroscopic: its Schwarschild radius should be much larger than the Planck scale. For this to happen, the parameters $g_sQ_i>>1$ for all $Q_i$.
This implies that although we have suppressed closed string 
interactions the interactions of the D-brane modes which have coupling constants $g_sQ_i$ are strong.
   
We have thus two distinct limits:

$\bullet$ $g_s Q_i<<1$. Here the bound state is point-like, but we can count states since we are dealing with weakly coupled gauge theory.

$\bullet$ $g_sQ_i>>1$. Here the gauge theory is strongly coupled and we cannot compute microscopically. In this region, the bound-state is a macroscopic black-hole.

Supersymmetry bridges the gap between the two regions. The states we will be counting will be unpaired BPS states. Thus, unbroken 
supersymmetry guarantees that the counting at weak coupling holds true at strong coupling.

To count at weak coupling we note that the effective two-dimensional theory
is a hyper-K\"ahler $\sigma$-model with central charge $c=6(Q_1Q+1)\simeq 6Q_1Q_5$ for large charges.
We need the density of states at level $Q_0$ and this is given by the Cardy formula in Conformal Field Theory: $\rho\sim exp[2\pi\sqrt{Q_0c/6}]$ which gives for the entropy 
\be
S_{\rm microscopic}=\log\rho=2\pi  \sqrt{Q_0Q_1Q_5}+\cdots=S_{\rm Bekenstein}+\cdots
\ee
where the ellipsis stands for subleading contributions.
The two results agree. The subleading contributions have been compared too
and agree. In gravity the correction comes from $R^4$ terms in the effective action.

This is the first example known where a microscopic counting of black-hole degrees of freedom agrees with the semiclassical, gravity result.
Similar agreement is found for more general extremal and near extremal black holes in five and four dimensions.

A further question concerns a more involved calculation: that of the Hawking radiation emission rate.  This rate, known as a grey-body factor (since it encodes also the interaction of the outgoing radiation with the gravitational field)
has a non-trivial dynamical content and it is not protected 
by supersymmetry in general. However, we have good reasons to believe that 
when supersymmetry is slightly broken (near extremality) the calculation is reliable at strong coupling \cite{das}.

\begin{figure}
\begin{center}
\vspace{0cm}  
\epsfig{file=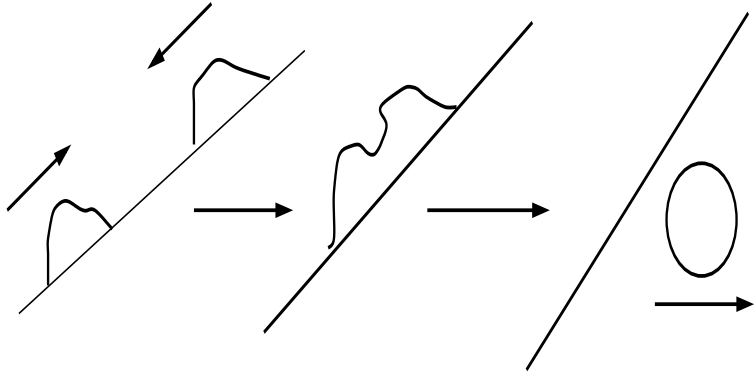,width=10cm,height=5cm}
\caption{The process of Hawking evaporation in the D-brane black-hole}
\label{fig13}
\end{center}
\end{figure}

In our five-dimensional example, near extremality means to add a small admixture of right-moving waves in the two-dimensional conformal field theory.
When a right and a left-moving wave scatter, closed strings can be produced: this is the Hawking radiation \cite{cama} (see figure \ref{fig13}).

Needless to say that the grey-body factors calculated from the D-brane approach 
agree with the gravitational calculation \cite{gr1}.

There are two open problems. Extend the above to black holes far away from the supersymmetric limit (Schwarschild for example).
And solve  the information paradox.
We should say that at weak coupling the description is manifestly unitary.
The question is: where the horizon is formed and whether there is an associated 
phase transition. The problem is open.

\section{ Gauge theory/gravity correspondence}
\setcounter{equation}{0}
 
We have discussed earlier in these lectures an intuitive picture of color confinement: electric flux is forced into thin flux tubes, that behave like strings with energy rising linearly with the distance and responsible for permanent confinement of quarks (that are attached at the ends of such flux tubes).
It is known since the early days of the $1/N_c$ expansion, \cite{largen}
that such a string description of the strong interactions becomes a good approximation when the number of colors $N_c\to \infty$.
In particular we would expect that different classes of large-$N_c$ gauge
theories to give rise to different effective string theories.

The running coupling of SU($N_c$) gauge theory is given by
\be
\mu{\partial\over \partial\mu}g_{YM}=-{11\over 3}N_c{g^3_{YM}\over 16\pi^2}+{\cal O}(g^5_{YM})
\ee
In order to have a regular expansion for the coupling we define the 
't Hooft coupling
\be
\lambda=g^2_{YM}N_c\sp \mu{\partial\over \partial\mu}\l=-{11\over 3}{\l^2\over 8\pi^2}+{\cal O}(\l^4)
\ee 
  
The standard large-$N_c$ limit is defined as $N_c\to \infty$ keeping $\l$ fixed and $small$.
To leading order, SU($N_c$) is indistinguishable from U$(N_c$), the propagator of the gauge field $A^{i\bar j}_{\mu}$ can be given as two lines, one for the fundamental index i and one for the anti-fundamental index $\bar j$ and thus
the typical perturbative diagrams automatically fatten up to become 
two-dimensional surfaces.
Moreover, the weighting factor for the diagrams scales a $N_c^{2-2g}$ where 
$g$ is the genus of the two-dimensional surface spanned by the fattened graph.
This is the first and important indication that the theory in this limit is described by some string theory although the nature of this string theory, despite many attempts over the years remained elusive.

The general expansion structure of (perturbative) observables is
\be
O(\l,N_c)=\sum_{g=0}^{\infty}~N_{c}^{2-2g}~C_{g}(\l)\sp C_{g}(\l)=\sum_{M=0}^{\infty}~C_{g,M}~\l^M\sp \l<<1
\ee
The would be string coupling constant is given by $g_s=1/N_c$.

This string cannot live in four flat dimensions. For one reason it is well
known that no string theory in flat space has Lorentz invariance apart from ten or twenty six dimensions.
As Polyakov \cite{poll} 
argued one needs at least an extra dimension in order to match symmetries (like the zig-zag symmetry) with the Wilson loop of 
the gauge theory.

The new idea in this direction is: A D-brane carries a gauge theory on its 
world-volume. This theory in a certain regime can reproduce gravitational effects (and vice versa).
By now this is not a complete surprise, since already in our discussion of
black hole entropy in the previous section, we have seen examples of this.
In fact, this correspondence was inspired by the black-hole investigations.
The crucial point here is two dual descriptions of a single object.
This object is a bound-state  of many branes.
At weak t'Hooft coupling, its physics is best described by weakly 
coupled gauge theory. At strong coupling the bound-state is macroscopic
and self-gravitating and can be well described by gravity.

We will look more carefully in the simplest example of this correspondence.
Consider a solitonic (gravitational) solution of type-IIB supergravity, the black D3-brane:
\be
ds^2={-f(r)dt^2+d\vec x^2\over \sqrt{H(r)}}+sqrt{H(r)}\left({dr^2\over f(r)}
+r^2d\Omega^2_5\right)
\label{metric}\ee
The three coordinates $\vec x$ and time describe the word-volume of the 
D3-brane while r and the five angles on $S^5$ parameterize the transverse space.
\be
H(r)=1+{L^4\over r^4}\sp f(r)=1-{r_0^4\over r^4}
\ee
There is also a non-zero spherically symmetric self-dual 
four-form $C_4$ \cite{hs}.

The position of the horizon is $r=r_0$.
$L^4=g_s N l_s^4$, where N is the charge (number) of the D3 brane.
Supergravity is a good approximation in this background when the curvature is
small compared to the string scale $L>>l_s$. This implies $g_sN>>1$.
On the other hand $g_s<<1$ so that gravitational loops are suppressed.
The limits are compatible when $N\to \infty$.

We will keep the ``distance" with units of energy $u=r/l_s^2$ fixed and we will take the $l_s\to 0$ limit in order to decouple unnecessary string modes.
This is the near-horizon limit of the black D3-brane.
\be
H(r)\to {g_sN\over l_s^4u^4}
\ee
and the metric becomes (we set $r_0=0$)
\be
ds^2=l_s^2\left[u^2(-dt^2+d\vec x^2)+{du^2\over u^2}+d\Omega_5^2\right]
\ee
which is the metric of $AdS_5\times S^5$.
This metric has the symmetry O(2,4)$\times$O(6) as well as maximal supersymmetry
(32 supercharges).
$AdS_5$ has a boundary at infinity $u=\infty$, that it is isomorphic to four-dimensional Minkowski space and can be reached at finite time from any point of the interior.

We will consider now the same object as a collection of N parallel D3-branes.
We have two kinds of excitations, open strings (fluctuations of the D3-branes
and closed strings,( bulk fluctuations).
The effective action will have the form
\be
S_{D3}=S_{\rm bulk}+S_{\rm Brane}+S_{\rm interaction}
\label{act}\ee
We will take the limit $l_s\to 0$ keeping dimensionless parameters fixed.
The bulk theory in this limit becomes free gravity (since gravity is IR free).
The same is also true for the interaction action that describes the interactions of brane and bulk fields.
Thus the only non-trivial interactions that are left over are the interactions
of $S_{\rm brane}$ namely those of N=4 U(N) super Yang-Mills.

In the previous supergravity description, as $l_s\to 0$ the excitations at $r\to \infty$  decouple from those near  the horizon. This is due to a potential barrier near the horizon that makes absorption cross sections to vanish with vanishing energy as $\s_{abs}\sim \omega^3~L^8$.
For the excitations near the horizon, the potential barrier keeps them from spreading out.

Moreover, as $l_s\to 0$ the fluctuations further away from the horizon
are described by free supergravity.
Matching the two descriptions we obtain the

{\bf Maldacena Conjecture} \cite{n}: N=4 D=4 SU(N) super Yang-Mills is dual to
type-IIB string theory on $AdS_5\times S^5$.
The gauge theory description is weakly coupled when $\l<<1$
while the supergravity description is insensitive to stringy data when 
$\l>>1$.
The symmetries of the supergravity theory match the conformal O(2,4) symmetry of Yang-Mills as well as its O(6) R-symmetry. The extra enhanced supersymmetry is due to conformal invariance.
It should be stressed however that for this correspondence $supersymmetry$ is not $important$.

The precise form of the correspondence states \cite{poll2} that 
quantum correlators in Yang-Mills match associated ``S-matrix elements" in
supergravity\footnote{Strictly speaking there are no S-matrix elements in AdS.
However at tree level we can define them using the usual procedure (equations of motion). However, they do not have the traditional interpretation in terms of scattering , but as we saw, they have to do with boundary correlators.}

Many strong coupling data of the gauge theory can be simply computed in the supergravity picture. A typical example are the Wilson loops\cite{wloo}, but 
also the particles (glueball spectrum \cite{glue}etc.)

Supersymmetry can be broken in two ways: turn on temperature in a higher dimension \cite{wit3} or find non-supersymmetric brane solutions of the 
effective equations \cite{sus1}.

Although this correspondence is a major step forward towards 
understanding gauge theory and gravity an important question still remains:
find the right QCD string!

\section{Conclusions and outlook}  

We have gone a long journey through some major theoretical developments of the past five years.
In the context of field theories major progress has been made towards understanding the strong coupling dynamics of supersymmetric theories.
The vacuum structure of N=1 gauge theories or the low energy effective action
of N=2 theories are some of the cornerstones of the effort.
A key ingredient in the above is electric-magnetic duality.
Supersymmetric theories naturally admit the concept that seems to capture 
some properties of the dynamics.

We have gone further and applied similar techniques based on dualities to supersymmetric string theory.
The outcomes are:

$\bullet$ We have learned that there is unique theory encompassing different looking  string theories.

$\bullet$ There is a most symmetric vacuum in eleven dimensions corresponding to a theory coined M-theory whose low energy limit is eleven-dimensional supergravity. M-theory unifies many apparently dual descriptions in lower dimensions.

$\bullet$ String theory contains many new objects, D-branes and other branes
that are essential for the consistency of the theory.

$\bullet$ D-branes provide a new and deep link between gauge theory and gravity.
They hint at quantization of space-time. They provide the microscopic degrees
of freedom responsible for black-hole entropy and might illuminate the puzzles 
of quantum gravity.

$\bullet$ In certain low energy limits they provide a link between gauge theory and gravity that leads to supergravity (or stringy) descriptions of gauge theories. The hope is that this will lead to a string theory for QCD.

There are many problems that are not yet solved. More relevant here is supersymmetry breaking for duality treatments of strong coupling problems.
Although there are some cases analyzed , it is not known how much of the non-perturbative treatments survive supersymmetry breaking.
The expectation is that for soft susy breaking and small susy breaking parameters, duality related non-perturbative techniques are still applicable.
This as well as strong breaking are open problems.

Another open problem is the search for the QCD string. Although the results so far are negative in the context of the gauge theory/gravity correspondence, there does not seem to be a reason forbidding its existence.
It maybe that the supergravity (or brane configuration) turns out to be too complicated. In any case, it is one of the  important problems of theoretical 
high energy physics.

Finally, applications of the above to the physics of the Standard Model and beyond  will be a concrete way to emphasize the value of these developments.

\vskip 3cm
\centerline{\Large\bf Acknowledgements}
\vskip 1cm

The author wishes to thank the organizers of the 99' Cargese Summer School on Particle Physics for hospitality and the students for creating a stimulating atmosphere.
This work was partially supported through a TMR contract
ERBFMRX-CT96-0090 of the European Union.


\newpage


\begin{thebibliography}{999}




\bibitem{is}
K.~Intriligator and N.~Seiberg,
{\it ``Lectures on supersymmetric gauge theories and electric-magnetic  duality,''}
Nucl.\ Phys.\ Proc.\ Suppl.\ {\bf 45BC} (1996) 1;
hep-th/9509066.



\bibitem{bil}
A.~Bilal,
{\it ``Duality in N=2 SUSY SU(2) Yang-Mills Theory: A pedagogical introduction to the work of Seiberg and Witten,''};
hep-th/9601007.

\bibitem{lerche1}
W. Lerche, {\it ``Introduction to Seiberg-Witten Theory and its Stringy Origin"},
 Nucl. Phys.  {\bf 55B} [Proc.Suppl.] (1997) 83; 
Fortsch. Phys. {\bf 45} (1997) 293; hep-th/9611190.



\bibitem{ag1}
L.~Alvarez-Gaume and S.F.~Hassan,
{\it ``Introduction to S-duality in N = 2 supersymmetric gauge theories: A  pedagogical review of the work of Seiberg and Witten,''}
Fortsch.\ Phys.\ {\bf 45} (1997) 159;
hep-th/9701069.



\bibitem{peskin}
M.E.~Peskin,
{\it ``Duality in supersymmetric Yang-Mills theory''}, 
hep-th/9702094.




\bibitem{dij}
R.~Dijkgraaf,
{\it ``Fields, strings and duality''},
hep-th/9703136.



\bibitem{shif}
M.~Shifman,
{\it ``Nonperturbative dynamics in supersymmetric gauge theories,''}
Prog.\ Part.\ Nucl.\ Phys.\ {\bf 39} (1997) 1;
hep-th/9704114.


\bibitem{ag2}
L.~Alvarez-Gaume and F.~Zamora,
{\it ``Duality in quantum field theory (and string theory),''},
 hep-th/9709180.


\bibitem{div} P. Di Vecchia, 
{\it ``Duality in N=2,4 Supersymmetric Gauge Theories"}, hep-th/9803026.


\bibitem{GSW} 
M. Green, J. Schwarz and E. Witten, {\em Superstring
Theory,
Vols I and II}, Cambridge University Press, 1987.

\bibitem{LT} 
D. L\"ust and S. Theisen, {\em Lectures in String
Theory},
Lecture Notes in Physics, 346, Springer Verlag, 1989.

\bibitem{LAG} L. Alvarez-Gaum\'e and M. Vazquez-Mozo, {\em Topics
in String Theory and Quantum Gravity}, in the 1992 Les Houches
School, Session LVII,
eds. B. Julia and J. Zinn-Justin, Elsevier Science Publishers, 1995.

\bibitem{polch} 
J. Polchinski, ``{\em What is String Theory?"}, hep-th/9411028.

\bibitem{oo}
H.~Ooguri and Z.~Yin,
{\it ``TASI lectures on perturbative string theories''},
hep-th/9612254.



\bibitem{book}
E.~Kiritsis,
{\it ``Introduction to superstring theory,''} Leuven University Press,  1998;
hep-th/9709062.

\bibitem{pol3}
J. Polchinski,
{\it ``String Theory"}, Cambridge University Press,1998. 





\bibitem{dkl} M. Duff, R. Khuri and J. Lu, 
{\it `` String Solitons"},
Phys. Rept. {\bf 259} (1995) 213; hep-th/9412184.

\bibitem{pcj} 
J. Polchinski, S. Chaudhuri and C. Johnson, {\it ``Notes on
D-branes"},  hep-th/9602052.

\bibitem{mal} 
J. Maldacena, {\it ``Black Holes in String Theory"}, PhD Thesis; 
hep-th/9607235. 

\bibitem{Polchinski:1996na}
J.~Polchinski,
{\it ``TASI lectures on D-branes''},
hep-th/9611050.


\bibitem{fl} 
S. F\"orste and J. Louis, {\it ``Duality in String
Theory"},
hep-th/9612192.

\bibitem{vafa} C. Vafa, 
{\it ``Lectures on Strings and Dualities"},
hep-th/9702201.

\bibitem{lec} 
E. Kiritsis, {\it ``Introduction to Non-Perturbative
String Theory"}, hep-th/9708130.

\bibitem{lerc} 
W. Lerche, {\it ``Recent Developments in String
Theory"}, hep-th/9710246.

\bibitem{peet} 
A. Peet, {\it `` The Bekenstein Formula and String Theory
(N-brane Theory)"}, hep-th/9712253.

\bibitem{wl} 
B. de Wit and J. Louis, {\it ``Supersymmetry and Dualities
in Various Dimensions"}, hep-th/9801132.

\bibitem{wt} 
W. Taylor, {\it ``Lectures on D-branes, Gauge Theory and
M(atrices)"}, hep-th/9801182.


\bibitem{senn} 
A. Sen, {\it ``An Introduction to Non-Perturbative String
Theory"}, hep-th/9802051.



\bibitem{bach}
C.P.~Bachas,
{\it ``Lectures on D-branes''}
hep-th/9806199.


\bibitem{aharony}
O.~Aharony, S.S.~Gubser, J.~Maldacena, H.~Ooguri and Y.~Oz,
{\it ``Large N field theories, string theory and gravity,''}
hep-th/9905111.





























\bibitem{hooft} G. 't Hooft, Nucl. Phys. {\bf B79} (1974) 276.

\bibitem{pol}
 A. Polyakov, JETP Lett. {\bf 20} (1974) 194.

\bibitem{jz}
 B. Julia and A. Zee, Phys. Rev. {\bf D11} (1975) 2227.


\bibitem{gon} P. Goddard, J. Nuyts and D. Olive Nucl. Phys. {\bf B125} (1977) 1.

\bibitem{mo}
C. Montonen and D. Olive, 
Phys. Lett. {\bf B72} (1977) 117.

\bibitem{witten}
E. Witten, Phys. Lett. {\bf B86} (1979) 283.

\bibitem{BW} J. Bagger and
J. Wess, ``{\em Supersymmetry and Supergravity}" Princeton Series in Physics, second edition, 1992.

\bibitem{hls}
R. Haag, J. Lopuszanski and M. Sohnius, Nucl. Phys. {\bf B88} (1975) 257.

\bibitem{n=4} M. Sohnius and P. West, Phys. Lett. 100B (1981) 245; \\
S.Mandelstam, Nucl. Phys. B213 (1983) 149;\\ 
P.S. Howe, K.S. Stelle and P.K. Townsend, Nucl. Phys. B214 (1983) 519; Nucl. Phys. B236 (1984) 125;\\
L. Brink, O. Lindgren and B. Nilsson, Nucl. Phys. B212 (1983) 401. 

\bibitem{n=2} P.Howe and K.Stelle and P.West, Phys. Lett. B 124B (1983) 55;\\
P. West, Proceedings of the 1983
Shelter Island II Conference on Quntum Field Theory and Fundamental
Problems of Physics, edited by R. Jackiw, N. Kuri , S. Weinberg and
E. Witten (M.I.T. Press).


\bibitem{sen}
A.~Sen,
Phys.\ Lett.\ {\bf B329} (1994) 217;
hep-th/9402032.



\bibitem{segal}
G.~Segal and A.~Selby,
Commun.\ Math.\ Phys.\ {\bf 177} (1996) 775.


\bibitem{vawi}
C.~Vafa and E.~Witten,
Nucl.\ Phys.\ {\bf B431} (1994) 3;
hep-th/9408074.



\bibitem{sw}
N.~Seiberg and E.~Witten,
Nucl.\ Phys.\ {\bf B426} (1994) 19;
hep-th/9407087\\
Nucl.\ Phys.\ {\bf B431} (1994) 484;
hep-th/9408099.


\bibitem{man}
S.~Mandelstam,
Phys.\ Lett.\ {\bf B53} (1975) 476;
Phys.\ Rev.\ {\bf D19} (1979) 2391;\\
G.~'t Hooft,
Nucl.\ Phys.\ {\bf B190} (1981) 455.



\bibitem{gauge1}
A.~Klemm, W.~Lerche, S.~Yankielowicz and S.~Theisen,
Phys.\ Lett.\ {\bf B344} (1995) 169;
hep-th/9411048;\\
P.C.~Argyres and A.E.~Faraggi,
Phys.\ Rev.\ Lett.\ {\bf 74} (1995) 3931;
hep-th/9411057.



\bibitem{sei}
N.~Seiberg and E.~Witten,
Nucl.\ Phys.\ {\bf B426} (1994) 19;
hep-th/9407087.



\bibitem{ak}
L.~Alvarez-Gaume, J.~Distler, C.~Kounnas and M.~Marino,
Int.\ J.\ Mod.\ Phys.\ {\bf A11} (1996) 4745; 
hep-th/9604004.





\bibitem{ven}
G. Veneziano, Nuovo Cimento, {\bf 57A} (1968) 190.



\bibitem{ss}
J.~Scherk and J.H.~Schwarz,
Nucl.\ Phys.\ {\bf B81} (1974) 118.



\bibitem{gs}
M.B.~Green and J.H.~Schwarz,
Phys.\ Lett.\ {\bf 149B} (1984) 117.


\bibitem{aw}
L.~Alvarez-Gaume and E.~Witten,
Nucl.\ Phys.\ {\bf B234} (1984) 269.



\bibitem{anto1}
I.~Antoniadis,
Phys.\ Lett.\ {\bf B246} (1990) 377.



\bibitem{dim1}
N.~Arkani-Hamed, S.~Dimopoulos and G.~Dvali,
Phys.\ Lett.\ {\bf B429} (1998) 263;
hep-ph/9803315;\\
I.~Antoniadis, N.~Arkani-Hamed, S.~Dimopoulos and G.~Dvali,
Phys.\ Lett.\ {\bf B436} (1998) 257;
hep-ph/9804398.


\bibitem{dim2}
N.~Arkani-Hamed, S.~Dimopoulos and G.~Dvali,
Phys.\ Rev.\ {\bf D59} (1999) 086004;
hep-ph/9807344.

\bibitem{k1}
E.~Kiritsis and C.~Kounnas,
Phys.\ Lett.\ {\bf B331} (1994) 51;
hep-th/9404092.



\bibitem{k2}
E.~Kiritsis and C.~Kounnas,
{\it ``String gravity and cosmology: Some new ideas''},
gr-qc/9509017.



\bibitem{gro}
D.J.~Gross, J.A.~Harvey, E.~Martinec and R.~Rohm,
Nucl.\ Phys.\ {\bf B256} (1985) 253; 
Nucl.\ Phys.\ {\bf B267} (1986) 75.

\bibitem{ht} 
C. Hull and P. Townsend, Nucl. Phys. {\bf B438} (1995)
109;
hep-th/9410167.


\bibitem{wit1}
E.~Witten,
Nucl.\ Phys.\ {\bf B443} (1995) 85;
hep-th/9503124;
{\it ``Some comments on string dynamics,''}
hep-th/9507121.




\bibitem{nep} 
R. Nepomechie, Phys. Rev. {\bf D31} (1985) 1921;\\ 
C. Teitelboim, Phys. Lett. {\bf B167} (1986) 69.

\bibitem{das}
S.R.~Das,
Nucl.\ Phys.\ Proc.\ Suppl.\ {\bf 68} (1998) 119;
hep-th/9709206;\\
E.~Kiritsis,
JHEP {\bf 10} (1999) 010;
hep-th/9906206.



\bibitem{cama}
C.G.~Callan and J.M.~Maldacena,
Nucl.\ Phys.\ {\bf B472} (1996) 591;
hep-th/9602043.


\bibitem{gr1}
S.R.~Das and S.D.~Mathur,
Nucl.\ Phys.\ {\bf B478} (1996) 561;
hep-th/9606185;\\
J.~Maldacena and A.~Strominger,
Phys.\ Rev.\ {\bf D55} (1997) 861;
hep-th/9609026.

\bibitem{largen}
G.~'t Hooft,
Nucl.\ Phys.\ {\bf B72} (1974) 461.



\bibitem{poll}
A.M.~Polyakov,
Nucl.\ Phys.\ Proc.\ Suppl.\ {\bf 68} (1998) 1;
hep-th/9711002.


\bibitem{hs}
G.T.~Horowitz and A.~Strominger,
Nucl.\ Phys.\ {\bf B360} (1991) 197.


\bibitem{n}
J.~Maldacena,
Adv.\ Theor.\ Math.\ Phys.\ {\bf 2} (1998) 231;
hep-th/9711200.


\bibitem{poll2}
S.S.~Gubser, I.R.~Klebanov and A.M.~Polyakov,
Phys.\ Lett.\ {\bf B428} (1998) 105;
hep-th/9802109;\\
E.~Witten,
Adv.\ Theor.\ Math.\ Phys.\ {\bf 2} (1998) 253;
hep-th/9802150.




\bibitem{wloo}
J.~Maldacena,
Phys.\ Rev.\ Lett.\ {\bf 80} (1998) 4859;
hep-th/9803002;\\
S.~Rey and J.~Yee,
hep-th/9803001.



\bibitem{glue}
C.~Csaki, H.~Ooguri, Y.~Oz and J.~Terning,
JHEP {\bf 01} (1999) 017;
hep-th/9806021.



\bibitem{wit3}
E.~Witten,
Adv.\ Theor.\ Math.\ Phys.\ {\bf 2} (1998) 505;
hep-th/9803131.




\bibitem{sus1}
I.R.~Klebanov and A.A.~Tseytlin,
Nucl.\ Phys.\ {\bf B546} (1999) 155;
hep-th/9811035.

















































\end{thebibliography}
\end{document}